\documentclass[10pt]{article}
\usepackage[english,activeacute]{babel}
\usepackage{natbib}
\usepackage{comment}
\usepackage{float}
\usepackage[hidelinks]{hyperref}
\usepackage{mathrsfs}
\usepackage{enumitem}
\usepackage[font={small,it}]{caption}
\usepackage{amsmath,amsfonts,amsthm,amssymb}
\usepackage{bm,rotating,multirow,dsfont,graphicx}
\usepackage[usenames, dvipsnames]{color}
\usepackage{url}
\usepackage{multicol}
\usepackage{multirow}
\usepackage[T1]{fontenc}
\usepackage{flafter}
\usepackage{appendix}
\usepackage{subfigure}
\usepackage{xcolor}
\makeatletter
\def\hlinewd#1{%
	\noalign{\ifnum0=`}\fi\hrule \@height #1 %
	\futurelet\reserved@a\@xhline}
\makeatother
\newcommand\simiid{\mathrel{\overset{\makebox[0pt]{\mbox{\normalfont\tiny\sffamily iid}}}{\sim}}}
\newcommand\simind{\mathrel{\overset{\makebox[0pt]{\mbox{\normalfont\tiny\sffamily ind}}}{\sim}}}

\def\@roman#1{\romannumeral #1}

\begin{document}

\def\spacingset#1{\renewcommand{\baselinestretch}{#1}\small\normalsize}\spacingset{1}

\title{Operationalizing Legislative Bodies: \\ A Methodological and Empirical Perspective}

\date{}
	
\author{
    Carolina Luque, Universidad Ean, Colombia\footnote{cluque2.d@universidadean.edu.co} \\ 
    Juan Sosa, Universidad Nacional, Colombia\footnote{jcsosam@unal.edu.co}
}	 

\maketitle

\begin{abstract} 
This manuscript extensively reviews applications, extensions, and models derived from the Bayesian ideal point estimator. 
We primarily focus our attention on studies conducted in the United States as well as Latin America. 
First, we provide a detailed description of the Bayesian ideal point estimator. 
Next, we propose a new taxonomy to synthesize and frame technical developments and applications associated with the estimator in the context of North American and Latin American governing bodies. 
The literature available in Latin America allows us to conclude that few legislatures in the region have been analyzed using the methodology under discussion. 
Also, we highlight those parliaments of Latin America embedded in democratic presidential systems as novel scenarios for operationalizing the electoral behavior of legislative bodies through nominal voting data. Our findings show some alternatives for future research. 
Finally, to fix ideas and illustrate the capabilities of the Bayesian ideal point estimator, we present an application involving the Colombian House of Representatives 2010–2014.
\end{abstract}


\spacingset{1.1} 


\section{Introduction}

The quadratic Bayesian spatial voting model (also known as the Bayesian ideal point estimator or IDEAL; \citealt{jackman2001multidimensional}, \citealt{clinton2004statistical}) is a valuable method for identifying latent (unobserved) traits of political actors from nominal voting data. 
This model is compatible with the nature of scientific research in the Social Sciences \citep{jackman2004bayesian,jackman2009bayesian}, and also, it is flexible and powerful enough for recognizing political preferences of voters \citep{carlin2008bayesian,clinton2009simulate}. 
Moreover, some authors point out that it is an essential tool for providing empirical evidence on different voting phenomena in modern Political Science \citep[e.g.,][]{clinton2004statistical,yu2021spatial, moser2019multiple}.

IDEAL and its derivatives have been widely used to analyze the electoral behavior of the United States Parliament. 
However, there are a limited number of applications in other contexts, such as Latin America. 
Therefore, there is a legitimate opportunity to characterize electoral phenomena of legislatures in Latin America \citep[e.g.,][] {tsai2020influence} by taking IDEAL as a reference framework due to its extensive usage in the North American Congress. We share this position with some authors, who see the theoretical and methodological developments of the legislative voting behavior of the United States Congress as a basis for advancing the quantitative understanding of parliamentary electoral conduct in scenarios with limited empirical evidence \citep{gamm2002legislatures}. 
Furthermore, the available literature shows that this path inspires adaptations and applications of IDEAL in different parliamentary electoral settings \citep[e.g.,][]{zucco2013legislative, mcdonnell2017formal, tsai2020influence,ribeiro2021legislative}.

This manuscript extensively reviews the available literature on models, extensions, and applications derived from IDEAL in North American and Latin American legislative settings. 
Our goal is to provide a comprehensive and up-to-date overview of statistical foundations, applications, and developments on the topic under consideration from 2001-2022. We present our findings by proposing a brand new taxonomy on the matter, considering the parliamentary voting phenomena underlying to use of the estimator. 
From a quantitative point of view, we focus on substantive hypotheses and methodological aspects that allow us to analyze legislative electoral behavior through roll call data.

To the best of our knowledge, no previous research discusses and synthesizes methodological aspects and applications of the estimator in the framework of parliamentary voting in both the United States and Latin America.
Our work is relevant to the scientific community for several reasons. 
First, it shows the fusion between political theory and statistical and computational principles as a fundamental element in implementing a spatial voting model in a particular legislative reality. 
Second, it reveals the limitations of the standard Bayesian ideal point estimator and shows them as an opportunity to promote methodological and applied research in the context of governing bodies.
Finally, it provides evidence of emerging scientific research niches in different legislative spaces. Mainly in Latin America, where legislatures rooted in presidential democratic systems have different dynamics than the North American parliament \citep[e.g.,][] {pereira2004cost,jones2005party,zucco2013legislative,ribeiro2021legislative}.

This document is structured as follows. 
Section \ref{sec:historia} shows spatial voting model terminology and the relevance of the ideal Bayesian point estimator through a historical review of studies in legislative committees. 
Section \ref{sec:Model} presents theoretical references related to the Bayesian ideal point estimator's specification, identification, and implementation. 
Section \ref{sec:NorthLatin} exhibits applications, adaptations, and extensions of the ideal point estimator in the context of North and Latin America. 
Section \ref{sec:illustration} illustrates the case of the Colombian House of Representatives 2010–2014.
Finally, Section \ref{sec:dis} synthesizes our main contributions and discusses some alternatives for future research.


\section{Voting models: Terminology and developments}\label{sec:historia}

The collective election is a phenomenon of interest in the political context because voters' preferences are not homogeneous \citep{krehbiel1988spatial,clinton2012using}. 
For example, in the case of deliberative bodies such as congress, legislators typically show different judgments about public policies under debate, which unequivocally leads to a collective conflict of perspectives. 
The \textit{spatial voting models} allow us to characterize this phenomenon, assuming that different subjects (legislators, judges, citizens, among others) adopt dissimilar preferences regarding choice alternatives (policies, proposals, motions, among others).

Most of the theoretical work on these models proposes normal or quadratic \textit{utility functions} to characterize the preferences of political actors \citep[to measure characteristics essential for understanding the causes and consequences of politics,][]{clinton2012using, carroll2013structure}. 
Such an assumption has its roots in the fields of Psychology and Mathematics. 
On the one hand, studies in psychology indicate that the response function of individuals when they judge similarities between stimuli or express their preferences is approximately normal \citep{poole2005spatial}. 
On the other hand, studies in Mathematics show that the normal distribution can be approximated by a quadratic expression through a second-order Taylor polynomial \citep{carroll2009comparing}. 
For an additional in-depth discussion about other utility structures in spatial voting models, see \cite{carroll2013structure} and \cite{tiemann2019shape}.

Assuming a \textit{quadratic utility function} in a spatial voting model is not only a convenient mathematical abstraction but also a standard formulation that describes the response of a subject facing a choice. 
Thus, each individual has an utility function centered around an \textit{ideal point}  \citep[optimal alternative,  see][]{clinton2012using} that represents his/hers preferences. 
Hence, the utility decreases as the \textit{distance} (metric defined on the reference space) between his/hers ideal point and the option of choice increases. 
In this way, the subjects vote for the alternatives closest to their ideal point according to a specific stochastic mechanism.

A spatial voting model fully characterizes the behavior of individuals facing an electoral process.
Additionally, it allows visualizing the \textit{political space} (reference space to represent the ideal points of individuals) in specific contexts such as that of a legislature.
The proximity of ideal points in the political space is evidence of similarities in legislators' voting records. 
It indicates the presence of \textit{latent traits} (e.g., ideological preferences) that underlie their voting behavior \citep{poole2005spatial}. 
Such an abstraction is advantageous because it allows for studying several issues, including the evolution of political parties, strategic voting, location of minorities, electoral interests and incentives, and relations between institutions.  
However, for this representation to make sense and be interpretable in a political framework, besides the statistical foundations of the model, it requires a deep understanding of the political system of the institutions under analysis \citep{poole1985spatial}.

\subsection{Formalizing electoral conduct of political actors}

Spatial voting models are involved in a quite extensive literature.
The foundational work of \citet{black1958theory} and \citet{downs1957economic} provide important advances for developing voting theory. 
Nevertheless, such work does not provide a formal mathematical structure to test political theories \citep{poole1985spatial}.
The first spatial voting models that rigorously operationalize electoral behavior appear almost a decade later, by identifying the position of the political space that candidates must adopt to win an electoral contest in a majority election \citep{davis1965mathematical,davis1970expository,hinich1970plurality}. 
These investigations formally introduce the concept of the utility function, pointing out its relevance in penalizing the discrepancy between the position of a voter and the voting alternatives as well as electorate preferences. 
However, these approaches do not consider a random component in voting procedures \citep{poole2005spatial}.

\newpage

The interest of researchers in proposing a rigorous mathematical structure allowing them to design measurement instruments to contrast political theory favored the development of several studies in subsequent decades.
For example, in the late 1970s, \citet{mckelvey1978competitive} investigated electoral behavior in committees through cooperative games \citep{aumann1964m}, which carries out the analysis in a new direction, but whose findings are not sufficient to support general patterns of strategic behavior \citep{arrow1990advances}.
Subsequently, based on agenda theory, \citet{romer1978political} and \citet{shepsle1979institutional} propose to explain the role that structures and procedures play in shaping the results of legislative voting. 
These authors provide initial results on the influence of the order of deliberation of motions on legislative voting behavior.
At about the same time, \cite{cahoon1978statistical} and \cite{wolters1978models} use multidimensional scaling methods to represent the responses of the voting process (dichotomous or polytomous) as low-dimensional continuous variables in \textit{Euclidean space}.
Subsequently, other studies investigate fundamental aspects of probabilistic voting behavior \citep{manski1977structure,coughlin1981electoral} and \textit{predictive/latent dimensions} \citep{enelow1984spatial}.

The probabilistic theory of electoral behavior considers votes as random variables whose behavior depends on \textit{stochastic utility functions}.
Such functions incorporate the distance between the voter's ideal point and the voting alternative and consider a random term \citep{mcfadden1976quantal,poole1984us} reconciling systematic spatial utility differences (intractable voting patterns) with probabilities to vote in favor of a given alternative.
Typically, \textit{random errors} are assumed to follow a normal or logistic distribution, leading to a probit or logit link function, respectively \citep{jackman2004bayesian,carroll2009comparing}. These two random mechanisms lead to similar one-dimensional results with no more than scale differences \citep{hahn2005probit,carroll2009comparing,lofland2017Assessing,luque2022bayesian}.

Predictive dimension methods frame voters and election objects in the same low-di\-men\-sio\-nal space to explain (dis)similarities between voters as a function of distances between points in the political space \citep{weisberg1970dimensions,poole1984us}.
Actors' voting patterns interpret \textit{dimensions of political space} in terms of issues, policies, or belief systems \citep{cahoon1978statistical, poole1985spatial, hinich1996ideology, hinich1997analytical}.
The number of dimensions of political space is a technical question. However, some authors have pointed out that it is common to identify low-dimensional spaces in the study of electoral behavior.
\cite{poole2005spatial} argues that only a few dimensions are required to capture the structure of the North American parliamentary voting behavior because legislators typically vote based on the underlying policies of the political group they represent (left-right, Republican-Democrat).
Furthermore, perceptual studies in psychology support the idea of a small number of dimensions underlying individual choice behavior, as people have a limited ability to perceive different objects \citep{miller1956magical} and a tendency to confront a group against another group \citep{ tajfel1981human}.

\subsection{Roll call data}

Spatial voting models based on data reduction techniques such as multidimensional scaling \citep{cahoon1978statistical}, factor analysis \citep{brazill2002factor}, and principal component analysis \citep{de2006principal} do not provide a methodological framework for modeling directly binary responses of individuals based on the parameters of interest. Then, these methods do not allow practitioners recovering voters' ideal points nor modeling the probability of voting either in favor of or against a given issue \citep{jackman2001multidimensional}.
The interest in explaining the individual behavior of deputies led to the development of parliamentary voting models in the 1980s \citep{poole1985spatial} focusing on estimating the ideal points of legislative actors and recovering their latent characteristics (political preferences). 
In essence, these are generalized linear models \citep{mccullagh2018generalized} that rely on iterative methods such as parametric bootstrap and Markov chain Monte Carlo \citep[e.g.,][]{clinton2004statistical, martin2002dynamic,lewis2004measuring,carroll2009comparing,carroll2009measuring} to estimate and infer political preferences from \textit{roll call data or nominal votes} \citep{poole1985spatial,poole1987analysis, poole2005spatial}.

By means of nominal votes legislators reveal their positions to other deputies and the general public \citep{mayhew1974congress}.
This political activity provides data that allows social scientists to estimate ideal points and retrieve the political preferences of political actors \citep{aleman2009comparing}.
Research using roll call data to learn about legislative decisions shows that these data are helpful to calculate correlations between parliamentary members through factor and cluster analysis methods, which are relevant to identifying aggregate patterns. However, its use to predict individual voting decisions is popular as well \citep[e.g.,][] {macrae1952relation,macrae1958dimensions,macrae1965method,vandoren1990can,poole2005spatial,clinton2012using,binding2022non}.
Most studies on nominal voting only consider binary data, i.e., data instances consisting of two alternatives, in favor of or against a motion, known as \textit{new alternative} and \textit{status quo}, respectively \citep{carroll2009comparing}.
Few studies in the literature consider abstentions \citep{thurner2000empirical,poole2005spatial,rosas2015no}. On the other hand, recent studies describe the implications and limitations of using roll call data to test political theory \citep[e.g.,][]{roberts2007statistical, ainsley2020roll}. Additionally, other research use this type of data with textual data to analyze legislative bodies from different data sources \citep[e.g.,][]{lauderdale2014scaling, kim2018estimating}.

In recent years, the use of roll call data to analyze parliamentary electoral behavior has increased substantially  \citep[e.g.,][]{clinton2004statistical,hix2005power,jones2005party,clinton2012using,zucco2013legislative,binding2022non,seabra2022beyond,rasmussen2022farmers,hansford2022estimating,grier2022campaign}.
In particular, the analysis of nominal votes of the legislative chambers of Latin American countries has become quite popular (see Table \ref{tab:t1}). Recent access to these data has made their analysis novel and relevant to regional legislative phenomena studies. Unfortunately, roll call data are not widely available in Latin America \citep{morgenstern2003patterns}. For example, \cite{zucco2013legislative} confirms the existence of a reduced number of roll calls to analyze parliamentary electoral behavior in the legislative chambers of Uruguay. Unlike the United States, in some countries in the region, roll call voting is not always the most frequent task \citep[see][]{jones2005party,ribeiro2021legislative}, and yet, it is becoming a common practice to facilitate access to this type of data \citep[see][]{mcdonnell2019congressbr}. Even when roll calls are available, these data are not systematized. Compiling them is a laborious task involving gathering information from external sources (local newspapers, websites, among others) \citep{morgenstern2003patterns, luque2021metodos}.

\begin{table}[ht]
\centering
\tiny 
\begin{tabular}{lllccccl}
\cline{4-7}
                                     &                 &                                    & \multicolumn{2}{c}{\textbf{Roll call data}} & \multicolumn{2}{c}{\textbf{Legislators}} &                                             \\ \hline
\multicolumn{1}{c}{\textbf{Country}} & \multicolumn{1}{c}{\textbf{Period}} & \multicolumn{1}{c}{\textbf{House}} & \textbf{Available}     & \textbf{Analyzed}     & \textbf{Available}    & \textbf{Analyzed}   & \multicolumn{1}{c}{\textbf{Reference}}     \\ \hline
Argentina           & 1989--2001     & Lower                              & 415                    &                    &                       &                  & \cite{morgenstern2003patterns}             \\
                                     & 1989--2003     & Lower                              & 473                    &                    &                       &                  & \cite{jones2005party}                      \\
                                     & 1993--1995     & Lower                              & 64                     &                    & 256                   &                  & \cite{rosas2008models}                     \\
                                     & 1989--2007     & Lower                              & 1193                   &                    &                       &                  & \cite{jones2009government}                 \\
                                     & 2002--2007     & Congress                               & 110                    & 31                 & 618                   & 356              & \cite{onuki2009political}                  \\
                                     & 2008--2009     & Lower                              & 251                    &                    &                       &                  & \cite{aleman2018disentangling}             \\
                                     & 1993--2017     & Lower                              &                        &                    &                       &                  & \cite{clerici2021legislative}              \\ \\
Brazil             & 1991--1998      & Lower                              & 623                    &                    &                       &                  & \cite{morgenstern2003patterns}             \\
                                     & 1989--2008     & Congress                               &                        &                    &                       &                  & \cite{mcdonnell2017formal} \\
                                     & 1989--2010     & Lower                              & 2332                   & 1806               &                       &                  & \cite{zucco2011distinguishing}             \\
                                     & 2003--2006      & Lower                              & 421                    &                    & 703                   &                  & \cite{tsai2020influence}                   \\ \\
Chile              & 2002--2006     & Lower                              & 157                    & 36                 & 120                   & 118              & \cite{onuki2009political}                  \\
                                     & 2004--2006     & Upper                              & 313                    & 118                &                       & 49               & \cite{aleman2008policy}                    \\ \\
Colombia            & 2006--2015     & Congress                               & 11600                  &              & 650                   &              & \cite{morales2021legislating}              \\
                                     & 2010--2014     & Upper                              & 417                    & 417                & 110                   & 91               & \cite{luque2022bayesian}                   \\ \\
Paraguay                             & 2003--2012       & Lower                              & 147                    &                    & 167                   &                  & \cite{ribeiro2021legislative}              \\ \\
Uruguay             & 1985--1994     & Congress                               & 63                     &                    &                       &                  & \cite{morgenstern2003patterns}             \\
                                     & 1985--2005     & Upper                              & 125                    &                    &                       &                  & \cite{zucco2013legislative}                \\ \cline{1-8} 
\end{tabular}
\caption{\textit{Available literature to analyze roll call data in Latin American legislatures. Some studies employ selection criteria for selecting either roll calls or legislators \citep[e.g.,][]{onuki2009political, zucco2011distinguishing, aleman2008policy, luque2022bayesian}. Recently, the nominal votes of Argentina, Brazil, and Colombia are available at \url{https://votaciones.hcdn.gob.ar/}, \url{https://bancodedadoslegislativos.com.br/}, and \url{https://congresovisible.uniandes.edu.co/}, respectively.}}
\label{tab:t1}
\end{table}

\cite{morgenstern2003patterns} manifests that some roll call voting records of the legislative chambers in Brazil, Argentina, Chile, and Uruguay have been available since the early 90s. For other countries in the region, these data have been made available later. 
For example, in the case of Colombia, parliamentary voting records have been collected since 2006 \citep{carroll2016unrealized}, and until very recently, they have been used to implement spatial voting models to retrieve political preferences of deputies \citep{luque2022bayesian}. 
In Latin America, the analysis of the lower house (House of Representatives) is more frequent compared to the upper house (Senate) (see Table \ref{tab:t1}). Few studies analyze the legislative behavior of the entire congress without discriminating by chambers. Under this scenario, we argue that Latin American legislatures are a young field for the empirical analysis of parliamentary electoral behavior based on roll call data.

\subsection{Operationalization of individual electoral behavior}

\citet{poole1985spatial} provide the building blocks that allow researchers to estimate ideal points and latent variables considering a political space from roll call data. Two popular models emerge from such foundational work: NOMINATE and IDEAL (technical details about the latter are available in Section \ref{sec:Model}).
In principle, both models are used to analyze the behavior of deputies in the Senate and House of Representatives, mainly from the United States. However, their application has been extended to other non-legislative contexts such as the United Nations \cite[e.g.,][]{voeten2000clashes,voeten2013data,bailey2017estimating,seabra2022beyond} and courtrooms \citep[e.g.,][]{martin2002dynamic, hansford2022estimating}.

NOMINATE is a frequentist spatial voting model based on an utility function with a random component \citep[see][for more details]{poole2001d,poole2005spatial,carroll2009measuring,caughey2016substance}. 
IDEAL is the Bayesian counterpart of NOMINATE \citep{clinton2009simulate}. The latter also incorporates random utility functions and extensively uses simulation-based approaches (e.g., Markov Chain Monte Carlo, MCMC) to estimate the legislators' ideal points  \citep{jackman2001multidimensional, clinton2004statistical}. 
There are comparative studies that highlight the similarities and differences between these tow models, as well as their advantages and disadvantages in modeling roll call data to retrieve political preferences from parliamentarians \citep[see][]{carroll2009comparing,clinton2009simulate}.
Research in this direction highlights the flexibility of IDEAL to test substantive hypotheses \citep{clinton2004statistical}. Furthermore, IDEAL is a more straightforward estimator than NOMINATE because it only requires the Bayes theorem for estimation and inference, making it a more generalizable method \citep{clinton2009simulate}. 
Also, IDEAL is an estimator that has a direct correspondence with the nature of scientific inquiry in the Social Sciences \citep[e.g.,][]{jackman2004bayesian,jackman2009bayesian}.

Thus, IDEAL has been extended in various ways. For example, \cite{martin2002dynamic} postulate a dynamic version that allows ideal points to change systematically over time. \cite{quinn2004bayesian} extends the model to handle non-binary data (including continuous data). \cite{treier2008democracy} formulate a variant of the model to measure the level of democracy in countries, while \cite{rosas2015no} do the same to analyze the behavior of the vote under informative abstentions. 
In another way, \cite{yu2020spherical} and \cite{yu2021spatial} develop spatial voting models based on non-Euclidean metrics. 
Also, there are applications of the model to study electoral behavior in legislatures other than the North American one. For example, \cite{zucco2013legislative} analyzes executive-legislative relations in Uruguay between 1985 and 2005 in the context of presidential coalitions. Furthermore, \cite{tsai2020influence} studies the Brazilian Chamber of Deputies between 2003 and 2006, highlighting the effect of political incentives on electoral behavior.

The literature makes it explicit that IDEAL is a dominant methodological and theoretical instrument in modern Political Science \citep{yu2020spherical,moser2019multiple}. Moreover, all the works presented above show that the analysis of the political preferences of legislators from nominal votes is an area of active research.
Therefore, IDEAL constitutes a fundamental model worth investigating.

\section{IDEAL: Bayesian ideal point estimator} \label{sec:Model}

\subsection{Modeling}

Roll call data arise when $n$ legislators vote on $m$ motions (such as bills or legislative initiatives). Thus, each legislator $i \in \{1, \ldots, n \}$ takes a position in favor of (yea) or against (nay) motion $j \in \{1, \ldots, m \}$. Hence, we let $y_{i,j} \in \{0,1\}$ be the vote cast by legislator $i$ on motion $j$, with $y_{i,j}=1$ if such a vote turns out in favor of the motion, and $y_{i,j}=0$ otherwise. These votes in favor of or against motion $j$ are assumed to be points in a $d$-dimensional Euclidean space, known as\textit{political space}, and are denoted by $\boldsymbol {\psi}_{j}$ and $\boldsymbol{\zeta}_{j}$, respectively.

It is assumed that all the legislators have a political preference, i.e., each legislator $i$ has a latent (unobserved) factor $\boldsymbol{\beta}_{i} \in \mathbb{R}^d$ known as \textit{ideal point}. Thus, decisions are made based on a quadratic utility function given by
$U_i(\boldsymbol{\psi}_{j})=-\parallel\boldsymbol{\psi}_{j}-\boldsymbol{\beta}_{i}\parallel^2+\eta_{i,j}$ and
$U_i(\boldsymbol{\zeta}_{j})=-\parallel\boldsymbol{\zeta}_{j}-\boldsymbol{\beta}_{i}\parallel^2+\upsilon_{i,j}$,
where $U_i(\boldsymbol{\psi}_{j})$ and $U_i(\boldsymbol{\zeta}_{j})$ are the corresponding profits associated with legislator $i$ for voting in favor of or against motion $j$, respectively, and $\eta_{i,j}$ and $\upsilon_{i,j}$ are independent random deviations resulting from the uncertainty involved during the decision processes, such that $\textsf{E}(\eta_{i,j} -\upsilon_{i,j})=0$ and $\textsf{Var} (\eta_{i,j} - \upsilon_{i,j}) = \sigma^2_{j}$.

Rational choice theory (e.g., \citealp{yu2019spherical}) states that under the previous setting, legislator $i$ votes in favor of motion $j$ if and only if $U_i(\boldsymbol{\psi}_{j})> U_i(\boldsymbol{\zeta}_{j})$.
Hence, assuming that $\eta_{i,j}-\upsilon_{i,j}$ is normally distributed, i.e., $(\eta_{i,j}-\upsilon_{i,j}) \simind \textsf{N}(0,\sigma^2_j)$, it follows that
$$
\textsf{Pr}(y_{i,j}=1 \mid \boldsymbol{\zeta}_{j},\boldsymbol{\psi}_{j}, \sigma_{j}, \boldsymbol{\beta}_{i}) = \textsf{Pr}(\epsilon_{i,j} < \mu_{j}+\boldsymbol{\alpha}_{j}^{\textsf{T}}\boldsymbol{\beta}_{i})
= \Phi(\mu_{j}+\boldsymbol{\alpha}_{j}^{\textsf{T}}\boldsymbol{\beta}_{i})
$$  
where $\epsilon_{i,j}=(\upsilon_{i,j}-\eta_{i,j})/ \sigma_j$ is the random component, 
$\mu_{j}=(\boldsymbol{\zeta}_{j}^{\textsf{T}} \boldsymbol{\zeta}_{j}-\boldsymbol{\psi}_{j}^{\textsf{T}}\boldsymbol{\psi}_{j})/ \sigma_{j}$ is the \textit{approval} parameter (representing the basal probability of a vote in favor of motion $j$), $\boldsymbol{\alpha}_{j}=2(\boldsymbol{\psi}_{j}-\boldsymbol{\zeta}_{j})/ \sigma_{j}$ is the \textit{discrimination} parameter (representing the effect that ideal points have upon the probability of a vote in favor of motion $j$), and $\Phi(\cdot)$ is the cumulative distribution function of the standard normal distribution.

The previous specification fully characterizes the probability of observing a positive vote since
$y_{i,j} \mid \mu_j,\boldsymbol{\alpha}_j,\boldsymbol{\beta}_i \simind \textsf{Bernoulli}( \Phi(\mu_j+\boldsymbol{\alpha}^{\textsf{T}}_{j}\boldsymbol{\beta}_i))$. The likelihood associated with such latent factor model is given by
\begin{equation*}
    p(\mathbf{Y} \mid \{\mu_j\},\{\boldsymbol{\alpha}_j\},\{\boldsymbol{\beta}_{i}\})=\prod_{i=1}^{n}\prod_{j=1}^{m} \Phi(\mu_{j}+\boldsymbol{\alpha}_j^{\textsf{T}}\boldsymbol{\beta}_{i})^{y_{i,j}}\left [ 1-\Phi(\mu_{j}+\boldsymbol{\alpha}_j^{\textsf{T}}\boldsymbol{\beta}_{i})\right]^{1-y_{i,j}}\,,
\end{equation*}
where $\mathbf{Y}=[y_ {i, j}]$ is a binary rectangular matrix of size $n \times m$.
Finally, in order to complete the specification of the model and carry out full Bayesian inference, it is required to specify a joint prior distribution on the parameter space. A computationally convenient alternative that works well in practice consists in letting
$(\mu_{j},\boldsymbol{\alpha}_j) \mid \boldsymbol{a}, \mathbf{A} \simiid \textsf{N}_{d+1}( \boldsymbol{a}, \mathbf{A})$
and
$\boldsymbol{\beta}_i \mid \boldsymbol{b}_{i}, \mathbf{B}_{i} \simind \textsf{N}_d(\boldsymbol{b}_{i}, \mathbf{B}_{i})$,
where $\boldsymbol{a}, \mathbf{A}, \boldsymbol{b}_i$, and $\mathbf{B}_i$ are the model hyperparameters (known fixed quantities). Of course, more complex hierarchical specifications are also possible.

\subsection{Identifiability}\label{sec:identi}

The model parameters in are not identifiable. Specifically, note that Euclidean distances among ideal points $\boldsymbol{\beta}_i$ and the voting alternatives $\boldsymbol{\psi}_{j}$ and $\boldsymbol{\zeta}_{j}$ remain invariant under any translation, rotation, or reflection of the political space. Such a geometric occurrence ensures that both discrimination parameters and ideal points are not distinguishable for any voting pattern $\mathbf{Y}$. For instance, consider a rotation of the political space through an $d \times d$ orthogonal matrix $\mathbf{Q}$, i.e., $\mathbf{Q}^{\textsf{T}} \mathbf{Q}=\mathbf{I}_d$. Then,
$(\mathbf{Q}\boldsymbol{\alpha}_j)^{\textsf{T}} (\mathbf{Q} \boldsymbol{\beta}_i) = \boldsymbol{\alpha}_j^{\textsf{T}} \boldsymbol{\beta}_i$, for all $i$ and all $j$, 
and therefore, 
$p(\mathbf{Y} \mid \{\mu_j\},\{\boldsymbol{\alpha}_j\},\{\boldsymbol{\beta}_{i}\}) = p(\mathbf{Y} \mid \{\mu_j\},\{\mathbf{Q}\boldsymbol{\alpha}_j\},\{\mathbf{Q}\boldsymbol{\beta}_{i}\})$.
Such a phenomenon is also typical of latent space models for social networks \citep[e.g.,][]{sosa2022latent}.

The lack of identifiability require us to establish parameter constraints. A popular alternative consists in imposing restrictions on the mean and variance of the ideal points. Specifically, in the same spirit of \cite{jackman2004bayesian} and \cite{lofland2017Assessing}, letting $\boldsymbol{b}_{i}=\boldsymbol{0}_d$ and $\mathbf{B}_{i}=\mathbf{I}_d$, for $i=1m,\ldots,n$, is useful to overcome translation and scale issues. Furthermore, it is convenient fixing the position of $d + 1$ legislator, known as \textit{anchor legislators}, with known (but distinctive!) political patterns in the political space, since it allows the model to differentiate legislative tendencies \citep{rivers2003identification,clinton2004statistical}.

\subsection{Prior elicitation and computation}\label{sec:hiperYcomputo}

Along the lines of \cite{clinton2004statistical}, we recommend to set $\boldsymbol{a}= \boldsymbol{0}_{(d + 1)}$ and $\mathbf{A} = \sigma^2\mathbf{I}_{(d + 1)}$ with $\sigma^2$ an arbitrarily large constant (e.g., $\sigma^2 = 25$) in order to assign a zero-centered non-informative prior distribution to $\mu_j$ and $\boldsymbol{\alpha}_j$, for $j=1,\ldots,m$, aiming to emulate roughly the behavior of a diffuse state of information.
This choice is quite reminiscent of the hyperparameter elicitation in a standard linear regression model when there is no need of informative or empirical alternatives, such as the unit information prior \citep{kass1996selection} or the $g$-prior \citep{zellner1986assessing}.
Finally, notice that it is not required to set a large variance a priori for the ideal points. Actually, all what is required is a prior notion of scale for this set of parameters.

\subsection{Posterior Inference}\label{sec:estim_infer}

Considering data from $n$ legislators on $m$ motions, we have $dn+m(d+1)$ unknown model parameters to estimate. Thus, the posterior distribution is framed in a high dimensional space, which makes it analytically intractable.
Even though other deterministic approaches recognized by their efficiency are available (e.g., variational approximations; \citealt{ormerod2010explaining}), we strongly suggest Markov Chain Monte Carlo algorithms (MCMC; e.g., \citealp{gamerman2006markov}) to approximate the posterior distribution.
In particular, by means of a Gibbs sampler, we are able to produce a sequence of dependent but approximately independent draws from the posterior distribution, through iterative sampling from the full conditional distributions of $\mu_{j}, \boldsymbol{\alpha}_{j}$ and $\boldsymbol{\beta}_{i}$. Hence, make it possible to compute point and interval estimates from the corresponding empirical distributions.

Following \cite{albert1993bayesian}, we recommend to increase the parameter space by introducing a set of auxiliary variables $\{z_{i,j}\}$ such that $y_{i,j}\mid z_{i,j} = 1$ if $z_{i,j} >0$, and $y_{i,j}\mid z_{i,j} = 0$ if  $z_{i,j} \leq 0$, where $z_{i,j} \mid \mu_{j},\boldsymbol{\alpha}_{j},\boldsymbol{\beta}_{i} \simind \textsf{N}(\mu_{j} + \boldsymbol{\alpha}_{j}^{\textsf{T}}\boldsymbol{\beta}_{i}, 1)$.
Notice that we obtain exactly the original Bernoulli model specified above when integrating the $z_{i,j}$ variables out. 
This model reformulation makes it easier to draw directly from the full conditional distributions of $\mu_j$, $\boldsymbol{\alpha}_j$, and $\boldsymbol{\beta}_i$.

Let $\mathbf{\Theta} = (\mu_1,\ldots,\mu_m,\boldsymbol{\alpha}_1,\ldots,\boldsymbol{\alpha}_m,\boldsymbol{\beta}_1,\ldots,\boldsymbol{\beta}_n)$ be the full set of model parameters. 
The posterior distribution of $\mathbf{\Theta}$ is
\begin{align*}
p(\mathbf{\Theta}\mid\mathbf{Y}) &\propto \prod_{i=1}^n\prod_{j=1}^m p(y_{i,j}\mid z_{i,j}) \times \prod_{i=1}^n\prod_{j=1}^m \textsf{N}(z_{i,j} \mid \mu_{j} + \boldsymbol{\alpha}_{j}^{\textsf{T}}\boldsymbol{\beta}_{i}, 1) \\ 
&\hspace{3.5cm}\times \prod_{j=1}^m \textsf{N}_{d+1}(\mu_j,\boldsymbol{\alpha}_j\mid\boldsymbol{a},\mathbf{A})
\times \prod_{i=1}^n \textsf{N}_{d}(\boldsymbol{\beta}_i\mid\boldsymbol{b}_i,\mathbf{B}_i)\,.
\end{align*}

Moreover, let $\phi^{(b)}$ denote the state of parameter $\phi$ in the $b$-th iteration of the Gibbs sampling algorithm, for $b = 1,\ldots,B$. Then, such an algorithm in this case is as follows:
\begin{enumerate}
    \item Choose a starting configuration for each model parameter, say $z_{i,j}^{(0)}$, $\mu_j^{(0)}$, $\boldsymbol{\alpha}_j^{(0)}$, and $\boldsymbol{\beta}_i^{(0)}$, for $i=1,\ldots,n$ and $j=1,\ldots,m$.
    \item Update $z_{i,j}^{(b-1)}$, $\mu_j^{(b-1)}$, $\boldsymbol{\alpha}_j^{(b-1)}$, and $\boldsymbol{\beta}_i^{(b-1)}$, for $i=1,\ldots,n$ and $j=1,\ldots,m$, cycling until convergence:
    
    \begin{enumerate}
    \item Sample $z_{i,j}^{(b)}$ from $p(z_{i,j}\mid\mu_{j}^{(b-1)}, \boldsymbol{\alpha}_{j}^{(b-1)}, \boldsymbol{\beta}_{i}^{(b-1)}, y_{i,j})$, where
    $$
    p(z_{i,j}\mid\mu_{j},\boldsymbol{\alpha}_{j},\boldsymbol{\beta}_{i},y_{i,j})
    =\left\{\begin{matrix}
            \textsf{TN}_{(0,+\infty)} (z_{i,j} \mid \mu_{j}+\boldsymbol{\alpha}_{j}^{\textsf{T}}\boldsymbol{\beta}_{i}, 1) & \text{if} & y_{i,j}=1\,,\\ \\
            \textsf{TN}_{(-\infty,0]}(z_{i,j} \mid \mu_{j}+\boldsymbol{\alpha}_{j}^{\textsf{T}}\boldsymbol{\beta}_{i}, 1) & \text{if} & y_{i,j}=0\,.
            \end{matrix}\right.
    $$
    \item Sample $(\mu_{j},\boldsymbol{\alpha}_j)^{(b)}$ from $p(\mu_{j},\boldsymbol{\alpha}_{j} \mid \{\boldsymbol{\beta}_i^{(b-1)}\}, \{z_{i,j}^{(b)}\})$, where
    $$
    p(\mu_{j},\boldsymbol{\alpha}_{j} \mid \{\boldsymbol{\beta}_i\}, \{z_{i,j}\}) = \textsf{N}_{d+1}(\mu_{j},\boldsymbol{\alpha}_{j}\mid\boldsymbol{c}_{j},\mathbf{C})\,,
    $$
    with
    $\boldsymbol{c}_{j} = (\mathbf{A}^{-1} + \mathbf{E}^{\textsf{T}}\mathbf{E})^{-1}\left( \mathbf{A}^{-1}\boldsymbol{a} + \mathbf{E}^{\textsf{T}}\boldsymbol{z}_{\bullet j} \right)$
    and
    $\mathbf{C} = (\mathbf{A}^{-1} + \mathbf{E}^{\textsf{T}}\mathbf{E})^{-1}$,
    where
    $\mathbf{E}$ is a rectangular matrix whose $i$-th row is $(1, \boldsymbol{\beta}_{i})$, and $\boldsymbol{z}_{\bullet j}=(z_{1,j},\ldots,z_{n,j})$.
    
    \item Sample $\boldsymbol{\beta}_i^{(b)}$ from $p(\boldsymbol{\beta}_i\mid\{\mu_{j}^{(b)}\},\{\boldsymbol{\alpha}_{j}^{(b)}\},\{z_{i,j}^{(b)}\})$, where
    $$
    p(\boldsymbol{\beta}_i\mid\{\mu_{j}\},\{\boldsymbol{\alpha}_{j}\},\{z_{i,j}\}) = 
    \textsf{N}_{d}(\boldsymbol{\beta}_i\mid\boldsymbol{d}_{i},\mathbf{D}_i)\,,
    $$
    with
    $\boldsymbol{d}_{i} = (\mathbf{B}_{i}^{-1} + \mathbf{F}^{\textsf{T}}\mathbf{F})^{-1} \left(\mathbf{B}_{i}^{-1}\boldsymbol{b}_{i} + \mathbf{F}^{\textsf{T}}(\boldsymbol{z}_{i\bullet}-\boldsymbol{\mu})\right)$ and $\mathbf{D}_i = (\mathbf{B}_{i}^{-1} + \mathbf{F}^{\textsf{T}}\mathbf{F})^{-1}$, where
    $\mathbf{F} = [\boldsymbol{\alpha}_1,\ldots,\boldsymbol{\alpha}_m]^{\textsf{T}}$, $\boldsymbol{z}_{i \bullet}=(z_{i,1},\ldots,z_{i,m})$, and $\boldsymbol{\mu}=(\mu_1,\ldots,\mu_m)$.
    
    \end{enumerate}
\end{enumerate}

\section{Ideal Bayesian point estimator: North and Latin American case}\label{sec:NorthLatin}

An extensive review of the literature shows that the \textit{ideal Bayesian point estimator} is essential to analyze parliamentary electoral behavior at any level (whether individual or collective). Its application allows political scientists to support conjectures about individuals' political preferences and the latent characteristics that underlie legislative decisions, taking into account political theory as well as the singularities of the parliament under analysis. Furthermore, the flexibility of the estimator makes it adaptable to provide empirical evidence on the relationship between policymakers, institutional arrangements, and legislative outcomes.

To present our findings, we propose a brand new taxonomy considering the type of parliamentary phenomena as well as the scope of the application. 
The taxonomy is composed of seven parts:  Political space dimension (Section \ref{sec:space}), pivot legislators identification (Section \ref{sec:pivote}), voting restricted to the agenda's nature (Section \ref{sec:agenda}), evolution and change in preferences of political actors (Section \ref{sec:EvoChange}), influence of national political leaders and groups (Section \ref{sec:influence}), strategic abstentions (Section \ref{sec:abstention}), and extremes voting together (Section \ref{sec:extremes}).

\subsection{Political space dimension}\label{sec:space}

The political space dimension is a popular topic in the Political Science literature \citep[e.g.,][]{potoski2000dimensional, jackman2001multidimensional, talbert2002setting, aldrich2014polarization, dougherty2014partisan, roberts2016dimensionality}. 
Research in this direction highlights the discrimination parameters \citep{jackman2001multidimensional}, the voting proposal typology \cite{moser2019multiple}, and the use of weighted Euclidean distances \citep{binding2022non} as crucial elements to analyze the dimension of political space. 
Thus, social scientists are interested in learning about \textit{the number of latent features to model parliamentarians' electoral behavior}, \textit{the nature of the retrieved dimensions}, \textit{identifiability in multidimensional models}, and \textit{additivity of legislators' preferences}, among others.

The dimension of the political space $d$, from a technical perspective, corresponds to the number of latent characteristics needed to model the voting behavior of legislators properly. 
Its choice translates into a model selection problem, seeking to achieve a balance between the goodness-of-fit and the complexity of the model \citep{moser2019multiple}. 
Several works consider methodological and epistemological discussions on the political space dimension as well as specific mechanisms for its choice \citep{benoit2012dimensionality,de2012struggle,jackman2001multidimensional,lofland2017Assessing,moser2019multiple}.

The dimension of the political space can be studied through the discrimination parameters $\boldsymbol{\alpha}_{1},\ldots,\boldsymbol{\alpha}_{m}$. 
These parameters allow analysts to assess the political space dimension in conjunction with the goodness-of-fit of the model, discern substantive content of the recovered dimensions, and also, contrast substantive hypotheses about the dimensions that underlie the political space \citep{jackman2001multidimensional,luque2022bayesian}.
Those motions with statistically significant discrimination parameters (that is, those whose credibility interval does not contain zero) provide evidence about the number of latent features required to explain voting patterns. 
For example, in the one-dimensional case, a considerable number of discrimination parameters indistinguishable from zero strongly suggests to embed  the model in higher dimensions \citep{jackman2001multidimensional}. 
Thus, inspecting the content of those voting lists associated with not significant discrimination parameters can reveal the qualitative character of additional dimensions to be considered. 
Defining informative prior states about discrimination parameters favors handling the complexities that arise when considering higher-dimensional models \citep{jackman2001multidimensional}.

Some authors point out that interpreting the discrimination parameters as factor loadings is quite useful to identify the substantive content of the retrieved dimensions \citep{jackman2001multidimensional}. However, some others argue that the dimension of the latent space is not necessarily aligned with the content of the bills presented for voting \citep{moser2019multiple}. In such a situation, interpreting the dimensions concerning substantive issues is inconsistent. These interpretive perspectives lead to discussions regarding a distinction between the basic space dimension and a thematic space \citep[see][for more details]{moser2019multiple,poole2007changing}.

In a recent study, \cite{moser2019multiple} indicates that standard spatial voting models, such as the Bayesian ideal point estimator, assume that the dimension of the political space is fixed and common. 
However, this assumption is a limitation for analyzing political actors' preferences in different electoral domains and identifying individual voting patterns. 
In this sense, this author proposes a methodology based also on aggregation principles for analyzing nominal data \citep[see][]{roberts2016dimensionality} that allows modelers determining latent characteristics common to the chamber and each legislator, taking into account different voting domains. 
In other words, this approach assumes that legislators can reveal different preferences depending on the nature of the vote (economy, security, among others). Furthermore, this framework assumes the existence of subsets of legislators whose voting patterns for certain groups of votes are not explainable through the linear combination of the latent traits recovered for the entire group.

The standard Bayesian ideal point estimator assumes that political actors' preferences in every dimension are additively separable. 
Then, the utility of an actor is given by the weighted sum of the deviations along the dimensions. 
In order to test this assumption, \cite{binding2022non} introduce a statistical model allowing non-separability across dimensions. 
These authors affirm that political actors' preferences can be characterized better by multiple non-separable dimensions rather than a single dimension or multiple independent dimensions.

Just a handful of studies in the Latin American context focus on providing empirical evidence to justify the choice of the political space dimension. 
Most studies assume one or two dimensions considering the political context of the legislatures under analysis \citep[e.g.,][]{zucco2013legislative, zucco2011distinguishing}. 
For example, on the one hand, \citet{rosas2005ideological} assures that legislative politics in most of the countries of Latin America is one-dimensional. On the other hand, \citet{zucco2011distinguishing} point out that the presence of religious, linguistic, or ethnic parties justifies the existence of a second ideological dimension. Thus, the presence of coalitions, electoral districts, and regional or provincial divisions, among others, indicate possible higher non-ideological dimensions \citep{zucco2011distinguishing,zucco2013legislative}.

Lastly, \cite{jones2005party} and \cite{luque2022bayesian}, inspired by  \cite{jackman2001multidimensional}, examine the dimension of political space through the analysis of discrimination parameters. 
The estimates of this parameters provide empirical evidence to ensure that there is a single dimension underlying the policy space in the case of the nominal votes of the Argentine House of Representatives 1989--2003 and Senate of the Republic of Colombia 2010--2014, respectively. 
These authors state that a one-dimensional model is enough to model properly such roll call data.

\subsection{Pivot legislators identification} \label{sec:pivote}

The identification of pivot legislators includes studies focused on the estimation and inference of both ideal points and auxiliary quantities obtained as a function of model parameters. 
Such quantities allow researchers to recognize those members of parliament whose position in the political space is considered as relevant in order to understand what happens within the legislature \citep{clinton2004statistical}. 
In this line of work, the interest relies on \textit{the identity and position of pivot legislators (also known as fundamental legislators), extremists, minorities, among others}.

The notion of pivot legislator is fundamental for some theories of parliamentary behavior that characterize and predict law formulation processes based on the position of these legislators in the political space \citep{clinton2004statistical}. 
In the North American context, such theories indicate that pivotal legislators are those whose vote is critical to the success or failure of the legislative process. 
In particular, the vote these deputies can guarantee the closure of a debate and define a majority vote in an extraordinary legislative situation \citep[see][]{krehbiel1998pivotal, clinton2004statistical}.

In this way, \cite{clinton2004statistical} exposes a methodology based on the standard Bayesian one-dimensional ideal point estimator, in order to discern the identity and spatial location of the deputies who play a fundamental role within the legislative body. 
The methodology corresponds to a sequential and iterative three-step scheme, namely, (i) sample the legislators' ideal points from their joint posterior distribution; (ii) order the sampled ideal points in ascending order, and (iii) observe which legislators occupy a particular pivot or order statistics of interest. 
After repeating this scheme an arbitrarily large number of times, we can identify those legislators who are more or less likely to occupy a particular ordered position.

In Latin America, studies in this direction are scarce. From a political perspective, social scientists are still determining whether the fundamental theory has significant applications in the Latin American context or not. 
Very recently, \cite{luque2022bayesian} identify legislators from the Colombian Senate 2010--2014 who are more likely to be in a conservative position or in the extremes of the political spectrum. 
Their findings are limited to individualizing deputies rather than making inferences about other quantities that can be derived from ideal point estimates. 
Nevertheless, more efforts need to be carried out in order to determine order statistics within parliament.

\subsection{Voting restricted to the agenda's nature} \label{sec:agenda}

Research in this direction highlights the importance of estimating parameters associated with voting alternatives \citep[see][]{clinton2001agenda}, say $\boldsymbol{\psi}_{j}$ and $\boldsymbol{\zeta}_{j}$, as a mechanism to investigate issues related to \textit{the behavior of the status quo in a specific period}, \textit{the positioning of new policies in particular moments and contexts}, and \textit{the dependency relationship between votes under a specific agenda}.

Standard political preference estimators, such as the standard Bayesian ideal point estimator, do not incorporate the sequential nature of the agenda, i.e., these models assume that the retrieved parameters are not affected by a reordering of the voting sequence. 
This assumption omits helpful information to locate parameters associated with voting alternatives and determine the political space dimension, which may lead to unsuitable ideal points estimates for certain political instances. 
Consequently, the corresponding findings could support misleading interpretations regarding the legislative theories tested with these models \citep{clinton2001agenda}.

In order to incorporate the agenda's sequential nature into roll call data analysis, \cite{clinton2001agenda} extend the standard Bayesian ideal point estimator to a constrained agenda model, where voting alternatives are tied to a status quo parameter. 
In formal terms and under the assumption of complete knowledge of the order of the agenda, the status quo of the voting list $j$, say $\boldsymbol{\zeta}_{j}$, is assumed to be associated with previous voting alternatives parameters. 
Thus, if $\boldsymbol{\psi}_{j-1}$ was approved, then the new status quo is $\boldsymbol{\zeta}_{j}=\boldsymbol{\psi}_{j-1}$. On the contrary, if $\boldsymbol{\psi}_{j-1}$ was negated, then the new status quo is $\boldsymbol{\zeta}_{j}=\boldsymbol {\zeta}_{j-1}$. 
Therefore, voting in favor of motion $j$ represents a movement in the political space, but voting against it, avoids such a movement.

In the North American context, the restricted model provides estimates of ideal points more consistent with deliberate policies than the standard estimator because it reveals the relationship between the status quo and the last approved proposal. 
However, the model needs to provide a framework for examining endogenous agenda formation and strategic voting within parliament. 
Additionally, the analysis does not provide evidence of the consistency about the restricted and unrestricted estimators \cite{clinton2001agenda}.

Although the proposal of these authors allows us to cover a little-studied legislative phenomenon, the Bayesian ideal point estimator considering the order of the agenda implies a price to pay in terms of parsimony. 
Unlike the standard model, this extension includes more parameters to estimate and more technical difficulties. To the best of our knowledge, there are no implementations of the constrained estimator in the Latin American context.

\subsection{Change in preferences of political actors} \label{sec:EvoChange}

In this line of work, research premises lie on \textit{the stability of the political preferences of legislators under particular circumstances of the legislative process} (e.g., change of party). Also, it is of interest to analyze the \textit{discrepancies exhibited by deputies in their voting patterns when faced with different voting domains}. In addition, the \textit{differences they reveal in their electoral behavior when they legislate in different institutional bodies (e.g., chambers and commissions)}, among others. All these questions lead to an essential technical discussion about the importance of \textit{establishing common latent scales} that allow contrasting the electoral behavior of political actors \citep[e.g.,][]{asmussen2016anchors,shor2010bridge,shor2011ideological}.

The latter task, establishing common latent scales, makes reference to a methodological aspect that implies analyzing \textit{bridges that allow ideal points to be scaled on a standard scale to ensure that the contrasts between institutions, periods, or other scenarios are compatible}. Thus, \cite{shor2010bridge} asserts that it is not valid to equate ideal point estimates obtained separately since each set of ideal points produces estimates on a different latent scale. Therefore, one alternative consists in identifying ``bridge actors'' (e.g., legislators with common voting records) to \textit{anchor} latent spaces that allow analysts to confront the electoral behavior of subjects not voting simultaneously.

The change of preferences has been considered from different perspectives. For example, \cite{martin2002dynamic} postulate a dynamic version of the standard Bayesian spatial voting model to assess the change and possible dependencies in political actors' preferences over time. 
Although these authors do not present results in the parliamentary context, their model is a reference for characterizing the dynamics of legislators' political preferences, since  their research describes different approaches to measure changes in such preferences over time. 
Additionally, they indicate the theoretical structure to incorporate dynamic linear models \citep{west2006bayesian} in the estimation of ideal points and the methodology to support the corresponding Bayesian inference.

The  dynamic model specification is analogous to the base model but taking into account indexing over time. In this spirit, the Bayesian ideal point estimator in its dynamic version is a dynamic linear model of the form
$y_{i,j,t}^{*} = \mu_{j}+\boldsymbol{\alpha}_{j}^{\textsf{T}}\boldsymbol{\beta}_{i,t}+\epsilon_{i,j,t}$, 
where $\boldsymbol{\beta}_{i,t}$ is the ideal point corresponding to legislator $i$ at time $t$, and $\epsilon_ {i,j,t}$ is the stochastic deviation due to the uncertainty associated with the voting process over time. 
The other parameters have the same connotation as in Section \ref{sec:Model}. 
This model differs from the base model mainly in the prior distributions of the ideal points. \cite{martin2002dynamic} propose a dynamic prior distribution given by
$\boldsymbol{\beta}_{i,t} \mid \boldsymbol{\beta}_{i, t-1}, \boldsymbol{\Delta}_{\beta_{i,t}} \stackrel{\text {ind}}{\sim} \textsf{N}(\boldsymbol{\beta}_{i, t-1}, \boldsymbol{\Delta}_{\beta_{i,t}})$,
where $\boldsymbol{\Delta}_{\beta_{i,t}}$ is a hyperparameter describing temporal evolution variation.

In the Latin American context, particularly in Brazil, we are aware of an approach that applies the one-dimensional standard dynamic Bayesian ideal point estimator to contrast the electoral behavior of deputies in the Chambers of Parliament 1989--2008. 
In particular, \cite{mcdonnell2017formal} establishes a common latent scale between chambers, taking as a bridge the joint votes of its members. 
In Brazil, representatives and senators occasionally vote sequentially in Congress. With this strategy, the author equates legislative behavior to an educational test where subjects act on the same policies almost simultaneously.

Another perspective to study change in political actors' preferences is assuming that the preferences of deputies are stable, at least in short periods, but they are affected by particular circumstances of the legislative process \citep{clinton2004statistical, moser2019multiple}. 
In this sense, \cite{clinton2004statistical} analyze the change of deputies' preferences in the context of party change through the standard one--dimensional Bayesian ideal point estimator (considering a liberal--conservative latent trait). 
This author states that legislators who change party affiliation in the exercise of their office are crucial to identifying the effect of the party on the voting behavior of members of the legislative body. 
At the moment of change, other determinants of the vote remain constant (e.g., the constituency and the affiliation of the other legislators, among others), and the new affiliation may reflect a variation in the ideal points of the legislators as a result of such a change. 
Naturally, all legislators can present a variation in the estimates of their ideal points after the change. 
However, the variation for deputies who change parties is greater than those who maintain their political affiliation. 
In this sense, it is possible to parameterize the ideal points as $\beta_{i,1}=\beta_{i,0}+\delta_i$, where $\beta_{i, 0}$ and $\beta_{i,1}$ are the ideal points of legislator $i$ before and after the change, respectively. Thus, the relative change, $\delta_i$, of the $i$-th legislator will make it possible to analyze the variation of the corresponding ideal point.

From this angle, there are two main critical issues in the quantitative investigation of change in preferences of political actors, namely, the precision and the compatibility of the estimates mentioned above \citep{mccarty2001hunt, clinton2004statistical}. Regarding precision, Bayesian simulation allows us to identify the inaccuracy that arises when dividing the data set into pre and post-change. Furthermore, it provides us with uncertainty assessments for all the model parameters, showing the drop in accuracy of the estimates after splitting the dataset around the match change. 
Now, regarding producing comparable estimates, \cite{clinton2004statistical} points out that the posterior standardization of the ideal points limits the problem. However, \cite{lofland2017Assessing} question this approach. These authors point out three methodological inconsistencies. First, comparing the hierarchical order of the legislators through the estimates made before and after the change of party ignores that locations of different legislators are not independent. One legislator increases his rank when another or others decrease theirs. Second, implementing models before and after the change produces non-comparable ideal point estimates. Standardized estimates of ideal points do not guarantee that they share the same latent scale. And third, the party-switching hypothesis must consider the lack of fit when testing multiple hypotheses.

Then, \cite{lofland2017Assessing} propose a hierarchical model to induce a common scale without splitting the vote set. Furthermore, they assume that not all legislators change their preferences. The deputies with constant political preferences throughout the period under analysis are the bridge between the political spaces \citep[see][]{shor2010bridge, shor2011ideological}. The model uses zero-inflated Gaussian priors to identify the bridge legislators that connect the arbitrary ideological scales and make them comparable. Additionally, this estimator allows addressing multiple comparison problems simultaneously \citep[see][]{scott2006exploration, scott2010bayes}, taking into account Bayes factors and posterior probabilities. Although the proposal is inspired by \cite{martin2002dynamic}, the main goal is not modeling the evolution of deputies' preferences over time, but to test hypotheses of party change at a specific moment in the legislative process.

Inspired by the work of \cite{lofland2017Assessing}, \cite{moser2019multiple} propose another extension of the ideal Bayesian point estimator. 
The formulation does not focus on party switching, but on variation in legislators' preferences when they vote in different domains or issues. 
The estimation of ideal points of political actors in different voting domains constitutes a version of change in parliamentarians' preferences little studied.
The extension of the model provides a methodological framework to carry out contrasts at both individual and group levels about similarities and discrepancies that deputies' revealed preferences present when they make decisions in different voting domains. 
These extensions propose estimating the identity of ``party legislators'' (bridge legislators), voters whose revealed preference remains constant for all lists. 
The approach of these authors does not assume prior knowledge of the identity regarding partisan voters.

These extensions of the ideal Bayesian point estimator also allow researchers to investigate the relationship between the political space and the thematic space through a clustering model \citep[see,][]{moser2019multiple}, under which legislators may have different preferences for each group of votes. 
The use of previous groupings reduces the number of different positions for a legislator and contributes to less uncertainty in the estimates of the model parameters. 
Unlike \cite{lofland2017Assessing}, \cite{moser2019multiple} generalizes the model to an arbitrary number of groups by introducing prior distributions that divide the set of votes into $K$ groups determined by the analyst. The generality  in the model formulation opens the door to potential applications mainly in Latin American legislatures, where there are no studies in this direction to the best of our knowledge.

\subsection{Influence of national political leaders and groups} \label{sec:influence}

An issue of interest to scholars of the North and Latin American Congress is \textit{the influence of party leaders or national political groups on the electoral behavior of legislators}. 
Studies in this line focus on the effect of political groups (parties, coalitions, among others) on electoral behavior and the influence of incentives on deputies' voting decisions. In other words, the research emphasizes voting patterns that are not the product of sincere voting behavior.

\cite{clinton2004statistical} provides a methodology to interpret and operationalize the influence of political groups on parliamentary voting behavior. 
Not considering this phenomenon leads to ideal points estimates absorbing an impact common to the members of a particular group and revealing a greater polarization (e.g., partisan) in the preferences of the deputies. 
In this direction, party influence is a plausible mechanism to generate extra utility incentives associated with a specific political group. 
Then, an extension of the utility function to model directly partisan incentives is  
\begin{equation*} \label{eqn:utilidadEjemploClinton}
    U_i(\psi_{j})=-(\psi_{j}-\beta_{i})^2+ \delta_{j}^\textsf{D}+\eta_{i,j} \hspace{1cm} \text{and} \hspace{1cm} U_i(\zeta_{j})=-(\zeta_{j}-\beta_{i})^2+\delta_{j}^\textsf{R}+\upsilon_{i,j}\,,
\end{equation*}
where $U_i(\psi_{j})$ and $U_i(\zeta_{j})$ are the corresponding profits associated with legislator $i$ for voting in favor of or against motion $j$, respectively, $\delta_{j}^{\textsf{D}}$ and $\delta_{j}^{\textsf{R}}$ are the incentives that the legislator $i$ receives for voting positively depending on their affiliation party, where \textsf{D} represents Democrats and \textsf{R} Republicans. Finally, $\eta_{i,j}$ and $\upsilon_{i,j}$ are random shocks product of the uncertainty associated with the voting processes, which are assumed independent and identically with logistic distribution. These utility functions lead to a linear model of the form
\begin{equation*} \label{regLatenteEjemploClinton}
    y_{i,j}^{*} = \mu_{j}+\alpha_{j}\beta_{i}+\delta_{j}D_{i}+\epsilon_{i,j}\,, 
\end{equation*}
where $\epsilon_{i,j}=(\upsilon_{i,j}-\eta_{i,j})/ \sigma_j$, $\mu_{j}=(\zeta_{j}^{2}-\psi_{j}^{2})/ \sigma_{j}$ and $\alpha_{j}=2(\psi_{j}-\zeta_{j})/ \sigma_{j}$, $D_{i}$ is a dummy variable that takes the value of 1 if the legislator $i$ is a Democrat, and 0 otherwise, and $\delta_j=\delta_{j}^{\textsf{D}}-\delta_{j}^{\textsf{R}}$ is the net difference of specific democratic party incentives. 
If $\delta_j>0$, then the net incentive is for Democrats that vote in favor of motion $j$, while if $\delta_j<0$, then the incentive is for Democrats that vote against motion $j$.

\cite{clinton2004statistical} declare that this extension is advantageous compared to other methodological proposals to study party influence. Unlike other mechanisms, it does not require a differential analysis of items \citep[e.g.,][]{wainer1993differential}, nor the use of two-stage procedures \citep[e.g.,][]{snyder2000estimating}. These methods are criticized due to the bias occurring during the estimation process \citep[e.g.,][]{mccarty2001hunt}. In this sense, the Bayesian approach improves several implementations of party influence models available in the literature. Furthermore, the model is quite general since it augments utility functions with party-specific incentives for each motion.

This extended estimator has the identification restrictions of the standard model but also introduces restrictions for the parameters $\delta_j$ (in the same direction as \citealt{snyder2000estimating}). 
Application of the model leads to the identification of voting lists subject to partisan incentives and the estimation of statistically significant party effects (the 95\% credibility interval of $\delta_{j}$ does not overlap with zero) for the US 105th Senate. 
\cite{clinton2004statistical} state that although there is evidence consistent with party influence in only a third of closed votes (roll call votes decided by margins closer to 35\% - 65\%), the magnitude of these incentives is large and politically consistent with the behavior of the deputies. 
Estimating party-specific incentives for each roll call makes it possible to determine party polarization more precisely.

In the same line of thought, \cite{richard2012sparse} adapt the standard Bayesian ideal point estimator using sparse factor models \citep[e.g.,][]{pati2014posterior, bernardo2003bayesian, carvalho2008high} to reveal partisan patterns in roll call votes of the US Senate. 
The proposal incorporates theoretical referents of the multivariate probit model \citep[e.g.,][]{chib1998analysis}, the Gaussian factor model \citep[e.g.,][]{murray2013bayesian}, and scattered mass points at zero \citep[e.g.,][]{castillo2015bayesian} to identify cases in which partisanship is not predictive of roll calls. 
This formulation makes it possible to characterize covariance patterns in multivariate binary data to establish which bills and legislators do not rely on latent factors of partisan nature. 
In this way, \cite{richard2012sparse} state that the sparse factor probit model is a crucial exploratory tool for analyzing high-dimensional correlated categorical data. 
These authors point out that, like the standard estimator, their method is adaptive and indicates that an interesting extension would consist in adding either spatial or temporal autocorrelation components to scores. 
The dependency would explain the partisanship of senators who serve in consecutive congresses or are spatially close (for example, belonging to the same district).

In the case of Latin America, findings in this regard have shown a broader literature. 
The influence of the executive, leaders, and political groups (leaders of the majority party, coalitions, delegations, among others) in the legislative vote is remarkable in the region. 
Studies in this direction have analyzed the phenomenon in question from different perspectives. The investigations have shown that legislatures in Latin America have a structure and dynamics that differ from the US parliament \citep{pereira2004theory}. 
Additionally,  the executive--legislative relationship \citep{pereira2004theory, zucco2013legislative}, the consolidation of government--opposition coalitions \citep{carey1998parties, carroll2016unrealized, tsai2020influence, zucco2011distinguishing}, multipartyism \citep{ames2002party, neto2002presidential, figueiredo2000presidential, pereira2004theory,tsai2020influence,zucco2013legislative}, political actors and internal subdivisions (committees) or external (constituencies, delegations, among others) to the legislative chambers \citep{jones2005party, cheibub2009political, pachon2016s, tsai2020influence, clerici2021legislative}, are determining factors of parliamentary electoral conduct. 
There are even authors who argue that, in contrast to the ideological component, these have a greater impact on the legislative voting process \citep[e.g.,][]{aleman2018disentangling, tsai2020influence}.

Thus, \cite{tsai2020influence} proposes adjusting the Bayesian ideal point estimator to incorporate non-ideological factors (such as membership in the ruling coalition) in the analysis of roll call data. 
This model offers empirical evidence regarding the influence of political groups on the legislative voting behavior of the Brazilian Chamber of Deputies from 2003--2006. 
This author states that the extension of the model is necessary since the Brazilian assembly is an example of a legislature that reflects electoral behavior mediated by political negotiation (strategies to access political resources) and ideological preferences. 
The standard ideal point estimates are not optimal since they need to distinguish between the impact of coalition dynamics and the effect of individual political preferences that underlie voting behavior. 
Other spatial voting models postulated under a similar argument are those proposed by \cite{zucco2009ideology} and \cite{zucco2011distinguishing}. 
In this latest study, the authors present a version of the standard ideal point estimator in two dimensions (left--right and government--opposition).

The modeling strategy assumes that the political problem underlying nominal voting has a one-dimensional ideological content (left--right) underlying different voting decisions to be made by legislators with the same political position. 
Then, the government--opposition relationship explains the discrepancy in the election. 
In this sense, \cite{tsai2020influence} models the executive's influence on legislative voting through the party affiliation of the deputies and the party membership in the ruling coalition. 
Thus, the author assumes that legislators whose party belongs to the alliance have additional incentives, $\delta>0$ and $\lambda>0$, if they vote in favor of the government proposals and against the opposition proposals, respectively.

Analogous to the formulation of \cite{clinton2004statistical}, the additional incentives, product of the government--opposition conflict, are incorporated into the utility functions
\begin{equation*} \label{eqn:utilidad1}
    U_i(\psi_{j})=-(\psi_{j}-\beta_{i})^2 + \delta \cdot \textsf{MGov}_{k[i]}\textsf{PGov}_{j}+\eta_{i,j}
\end{equation*}
and
\begin{equation*} \label{eqn:utilidad2}
    U_i(\zeta_{j})=-(\zeta_{j}-\beta_{i})^2+\lambda \cdot \textsf{MGov}_{k[i]}\textsf{POpp}_{j}+\upsilon_{i,j}\,,
\end{equation*}
where $U_i(\psi_{j})$ and $U_i(\zeta_{j})$ are the corresponding profits associated with legislator $i$ for voting in favor of or against motion $j$, respectively, $\eta_{i,j}$ and $\upsilon_{i,j}$ are random shocks product of the uncertainty associated with the voting processes, which are assumed independent and identically distributed. 
Additionally, $k[i]$ denotes the party affiliation of the legislator $i$, in a way that $\textsf{MGov}_{k[i]}$ corresponds to a binary indicator that takes the value of 1 when legislator $i$ belongs to party $k$ of the governing coalition, and otherwise, it takes the value of 0. 
Finally, $\textsf{PGov}_j$ and $\textsf{POpp}_j$ are dummy variables that take the value of 1 when proposal $j$ comes from the government and the opposition, respectively, otherwise, they take the value of 0.

\newpage

The incentives $\delta$ and $\lambda$ directly affect the parameter intercept $\mu_j$ in the regression model 
$$\mu_j=\gamma_{1}+\gamma_{2}\textsf{PGov}_{j}+\nu_{j}\,,$$
where $\gamma_{1}$ represents the reference probability of granting a positive vote when the opposition casts the proposal and $\nu_{j}\simiid{\textsf{N}}(0,\sigma_{\nu}^2)$.
Thus, $\gamma_{2}>0$ indicates that the government--opposition conflict increases the probability of voting in favor when proposal $j$ is governmental, whereas $\gamma_{2}<0$ indicates that the coalition dynamic does not positively affect the government direction. 
The specification includes conventional semiconjugate prior distributions of the forms
$\sigma_{\nu}^2 \sim \textsf{GI}\left(a_0/2,b_0/2\right)$, 
$\gamma_{1} \sim \textsf{N}(\gamma_{0}, \sigma_{\nu}^2)$, and $\gamma_{2} \sim \textsf{N}(\gamma_{0}, \sigma_{\nu}^2)$, where $a_0$, $b_0$, and $\gamma_0$ are the model hyperparameters.

Additionally, the model incorporates the party affiliation through informative priors on the ideal points. The aim behind this approach is consists in evaluating the variation between parties and analyzing the effect of the ruling coalition on the legislative vote through a party base. 
Thus, in the same spirit of \cite{zucco2011distinguishing}, the ideal point corresponding associated with $i$ comes from a distribution specific to his political party, centered on the party mean $\mu_{\beta_{k[i]}}$ with variance $\sigma_{\beta}^2$, i.e.,   
$\beta_i\mid\mu_{\beta_{k[i]}},\sigma^2_\beta {\sim}{\textsf{N}}(\mu_{\beta_{k[i]}},\sigma_{\beta}^2)$. 
Unlike other authors, \cite{tsai2020influence} proposes priors hierarchies on party means. 
In this sense, he suggests that $\mu_{\beta_{1}} \mid \mu_{\beta_{2}}  {\sim} \textsf{U} (-3,\mu_{\beta_{2}})$, $\mu_{\beta_{k}} \mid \mu_{\beta_{k-1}},\mu_{\beta_{k+1}} {\sim} \textsf{U} (\mu_{\beta_{k-1}},\mu_{\beta_{k+1}})$,
and $\mu_{\beta_{K}} \mid \mu_{\beta_{K-1}} {\sim} \textsf{U} (\mu_{\beta_{K-1}},3)$, for $k=2,\cdots, K-1$.

In this way, $\mu_{\beta_{k}}$ is subject to a prior ordering restriction, through which political parties are organized from left to right, taking into account positions that other authors have identified \citep[e.g.,][]{zucco2009ideology, zucco2011distinguishing}. Additionally, considering the pre-established ordering, two parties with different ideologies are selected to anchor the extremes of the scale. Thus, this specification allows the problem of rotational invariance to be solved directly since it makes the left and right parties stand on the corresponding side of the underlying ideological scale. On the other hand,  electing deputies from ``anchor parties'' with different ideologies but belonging to the same field (government--opposition, selection based on \citep{zucco2011distinguishing}) ensures that the underlying ideological dimension and the government--opposition conflict do not overlap.
Lastly, the sign of the discrimination parameters $\alpha_{j}$ reveals the ideological status of the proposal $j$. The prior distributions for these parameters are 
$\alpha_{j} {\simind}{\textsf{N}}(-2,1)\boldsymbol{I}(\alpha_{j}<0)$, for $j=1,\ldots, m_{0}$, and
$\alpha_{j} {\simiid}{\textsf{N}}(0,1)$, for $j=m_{0},\ldots, m$,
where $\boldsymbol{I}(\cdot)$ denotes the indicator function. Model parameters are standardized in each iteration of the MCMC algorithm for solving scale invariance problems \citep{tsai2020influence}.

The model results regarding the Chamber of Deputies of Brazil reveal that legislators with membership in the government coalition are more likely to support the executive's proposals regardless of their political ideology. Empirical evidence confirms that parties influence legislative votes. In addition, estimating the ideal points associated with the 11 parties in the ideological scale (left--right) reveals a partisan order similar to previous studies \citep[e.g.,][]{zucco2011distinguishing}.

On the other hand, \cite{jones2005party} implement the standard estimator to roll call data from the Chamber of Deputies of Argentina 1989--2003. 
The estimated ideal points help build an indicator of homogeneity of provincial legislative behavior to analyze the influence of governors on the legislative behavior of their regional partners in the lower house. 
In turn, such a metric is considered as the response variable of a linear regression model that evaluates the governor's influence on parliamentary electoral behavior using the characteristics of the governor as covariates (e.g., belonging to the government). 
The results do not reveal provincial effects on the legislative behavior of deputies, nor do they reflect a significant relationship between the characteristics of the governor and the estimated measure of homogeneity. \cite{zucco2013legislative} also uses the estimated ideal points to build an index of legislative performance. 
This measure is the response variable of a Tobit model \citep{amemiya1984tobit} to explain the influence of political groups or actors on parliamentary voting behavior through covariates such as the average ideological distance of the faction from the president as well as the proportion of nominal passes in which the faction held a ministerial position. 
The pattern revealed by the estimated ideal points indicates that the behavior of legislators from the same fraction is alike. In addition, the ideological distance influences the legislative behavior indicator. Finally, legislators whose factions hold ministerial positions behave more pro-government than their ideology might predict.

Finally, \cite{zucco2013legislative}
also implements the standard one-dimensional estimator to roll call data from the Uruguayan Senate 1985--2005. A low number of votes (see Table) and a highly institutionalized party system lead to a distribution of ideal points in the political space that does not follow an ideological pattern. Therefore, the assumed dimension cannot be interpreted ideologically, but rather in terms of the executive--legislative relationship.

\subsection{Strategic abstentions} \label{sec:abstention}

Research in the North American context \citep[see][]{rodriguez2015measuring, rosas2015no} have proposed to extend the standard Bayesian ideal point estimator to study the phenomenon of strategic abstentions (intentional not voting, absenteeism from the room, not vote recording, among others; a theoretical explanation in this regard is given in \citealt{cohen1991vote, kromer2005determinants, voeten2000clashes}). 
Abstention from voting is a characteristic of the electoral behavior of deliberative bodies usually ignored without empirical support \citep{rodriguez2015measuring, rosas2015no}. 
The matter of interest associated with this issue  are \textit{the identity of legislators who frequently participate in strategic abstentions}, \textit{the behavior of the abstention rates of groups or political leaders over time or under specific conditions}, and \textit{the types of bills for which legislators locations in the political spectrum are key as abstention factors}, among others.

The standard Bayesian ideal point estimator typically ignores missing values and assumes that the random process generating these missing data is irrelevant for estimating the ideal points \citep{rosas2015no}. However, abstentions reflect legislators' preferences when faced with the dilemma of not affecting others (e.g., party leaders or voters) during the competition \citep{rodriguez2015measuring, rosas2008models}. Consequently, \cite{rosas2015no} point out that the decision to model the lack of response should be based on theoretical foundations and not on goodness-of-fit measures since the strategic nature of abstentions is specific to the context of each legislature.

The extensions of the estimator that address the phenomenon of abstentions show two main approaches: first, moving from a binary model to an ordinal model including the non-response as an intermediate category between the positive and negative vote \citep{rodriguez2015measuring}, and second, keep a binary model but consider the joint probability of both positive vote and vote registration \citep{rosas2015no}. These estimators characterize the mechanism that drives the presence of missing values in voting lists. Moreover, its implementation allows modeling the choice of non-response and voting simultaneously, as well as evaluating the impact of factors (e.g., as ideology) on the probability of non-response and vice versa \citep[see][]{rodriguez2015measuring, rosas2015no}.

\cite{rosas2008models} present an extension of the canonical ideal point estimator (standard model assuming completely random abstentions, also called ignorant abstentions model) to analyze the processes of abstention and voting simultaneously. They illustrate their proposal by analyzing roll call data from the Federal Congress of Argentina 1993--1995. These data favor the study due to the high rates of abstention as well as lack of recording (32\% for this period).

The one-dimensional bivariate probit model admits two sources of information: (i) the voting choice (1 = \textsf{Yes}, 0 = \textsf{No}) and (ii) the vote registration indicator (1 = \textsf{No registration} and 0 = \textsf{Registration}) of the legislators. Then, $y_{1,i,j}^{*}$ and $ y_{2,i,j}^{*}$ are linear predictors driving the probability of positive vote and abstention for legislator $i$ in motion $j$, respectively. These predictors are
$y_{1,i,j}^{*} = \mu_{j}+\alpha_{j}\beta_{i}+\gamma_jc_i+\epsilon_{1,i,j}$, with $\epsilon_{1,i,j} \simiid{\textsf{N}}(0,1)$,
and
$y_{2,i,j}^{*} = \eta_{j}+\delta_{j}c_{i}+\epsilon_{2,i,j}$, with $\epsilon_{2,i,j} \simiid\textsf{N}(0,1)$.
The former expression is the canonical latent model with an additional predictor $c_i$ that represents the latent propensity to abstention of legislator $i$ in motion $j$, and $\gamma_j$ is the effect of such propensity on the probability of a positive vote on motion $j$. 
These parameters are incorporated into the model in order to allow the abstention process to be informative of the result of a vote. 
On the other hand, the latter expression characterizes the mechanism that generates abstention, i.e., the probability that legislator $i$ abstains when considering vote $j$, with $\delta_j$ y $\eta_j$ are specific parameters associated with voting list $j$.

\cite{rosas2008models} define informative prior distributions for the parameters of the model under the assumption of the influence of the provincial delegations of the parties in the electoral behavior of the deputies \citep[see][]{jones2005provincial}. 
They presume that the propensities to the abstain are grouped within such delegations as they define a common prior for the members of the same delegation. 
Although there are 20 parties with representation in congress during the period of interest, the results only provide information on the voting process and abstention of the majority and opposition parties. 
The proposed model presents an adequate performance compared to the model that assumes ignorable abstentions. However, the number of roll call data for the analysis is scarce, and the information retrieved through them is insufficient  (see Table \ref{tab:t1}) to generalize the results. Nevertheless, these findings reveal information about voting lists and bills that provide the highest abstention rates and discriminate better between legislators.

\subsection{Extremes voting together} \label{sec:extremes}

Very recently, spherical latent factor Bayesian models emerged for binary and ordinal multivariate data \citep[see][]{yu2021spatial,yu2019spherical,yu2020spherical}. 
Although this kind of model is similar to the ideal Bayesian point estimator, it is not an extension. These models constitute a more general class, and the standard estimator is a limiting case of the new family of models \citep{yu2019spherical}.

\cite{yu2021spatial} mention that most spatial voting models assume ideal points embedded within a (possibly multidimensional) Euclidean political space. However, there are situations where Euclidean geometry does not explain unusual voting behavior. For example, the phenomenon where members from opposite ends of the ideological spectrum reveal similar preferences by voting against the rest of the legislature. In this case, neither increasing the dimensionality of the latent space nor performing linear transformations allow the classical estimator to characterize the phenomenon of ``extremes voting together'' adequately. As a result, the standard ideal point estimator does not perform well in the mentioned case since it exhibits the extreme deputies as conservative individuals. Instead, it is a better fit when nominal vote data comes from political systems where parties are relatively unified \citep{yu2021spatial}. So then, this new class of models explores \textit{unconventional voting patterns evaluating the geometry that underlies the political space}.

Bayesian spatial voting models based on spherical geometries differ from the classical approach in their specification and implementation \citep[see][]{yu2020spherical}. In particular, the model parameters, $\boldsymbol{\beta}_{i}$, $\boldsymbol{\psi}_{j}$ and $\boldsymbol{\zeta}_{j}$, are defined on a Riemann manifold $\mathcal{D}$ \citep[see][]{lee2018introduction}, and the Euclidean distance is replaced by the geodesic distance $\rho$ in $\mathcal{D}$. The distance between two points is defined on a unit $(d+1)$-dimensional hypersphere. This distance corresponds to the smallest angle formed by the projections of the points through the origin. The case $d=1$ leads to a spatial voting model in a circular space with utility functions given by
\begin{equation*} \label{eqn:utilidadEsferica}
    U_i(\psi_{j})=-\rho{(\boldsymbol{\psi}_{j},\boldsymbol{\beta}_{i})}^2+ \eta_{i,j} \qquad \text{and} \qquad U_i(\zeta_{j})=-\rho{(\boldsymbol{\zeta}_{j},\boldsymbol{\beta}_{i})}^2+\upsilon_{i,j}\,,
\end{equation*}
where $\rho(a,b)=\arccos{(\cos{(a- b)})} \in [0, \pi]$ is the geodetic distance that establishes the smallest angle between $a$ and $b$. Thus, $\beta_{i}$, $\psi_{j}$, and $\zeta_{j}$ $\in [- \pi, \pi]$ are interpreted as angular positions in an unit circle. The specification of the utility function leads to substantial changes in terms of the choice of the link function, the prior distributions of the model parameters, the identification, and interpretation of the model, and its computational implementation \citep[see details in][]{yu2019spherical,yu2020spherical,yu2021spatial}.

The circular version of this family of models was applied to 1988--2019 United States House of Representatives roll call data \citep[see][]{yu2021spatial}. The results reveal that the circular voting model explains modern voting patterns better than traditional Euclidean models. On the other hand, the circular variance as a measure of sphericity (extreme voting behavior) helps analyze the temporal evolution of this voting phenomenon.

Spherical Bayesian spatial voting models have yet to be widely applied. To the best of our knowledge, there is still a need for applications in the Latin American legislatures' context. Existing implementations are novel and provide empirical evidence of a recent phenomenon. Furthermore, these models suggest uninvestigated methodological paths. For example, work has yet to be reported in the literature illustrating how to incorporate and interpret strategic abstentions in this model. The applications discussed so far assume that missing data result from mere chance. Furthermore, there is an open path to explore suitable priors in high-dimensional spheres contexts \citep{yu2021spatial}.

\section{Illustration} \label{sec:illustration}

In order to fix ideas and provide a brand new case study, we focus on the Colombian House of Representatives 2010--2014. This chamber's plenary roll call votes have yet to be previously studied using some of the ideas under review. Therefore, we take advantage of the novelty of this data set to broaden the discussion on the nature of the dimension of the political space of the Colombian parliament, particularly the lower house. We have available 626 roll calls deliberated by 181 deputies. However, we omitted 31 legislators for participating in less than $95\%$ of the votes, which does not imply eliminating any political group. In addition, we eliminated 66 unanimous roll calls since these are not relevant to recover information about the latent traits that underlie the political decisions of parliamentarians \citep{jackman2001multidimensional,luque2022bayesian}.

Four political groups make up the House of Representatives 2010--2014, the coalition of government, independents, minorities, and opposition. Four of the fourteen political parties in the legislative exercise constitute the coalition, Conservador Colombiano (CC, $21.33\%$), Liberal Colombiano (LC, $22.67\%$), Cambio Radical (CR, $10\%$), and Social de Unidad Nacional (PU, $28.67\%$). The parties openly declared as independent are Alianza Verde (PAV, $2\%$), Integración Nacional (PIN, $6.67\%$), and MIRA ($0.67\%$). Five of the six minority parties have a seat in the chamber (each with $0.67\%$ representation). These are Alas Equipo Colombia (ALAS), Alianza Social Indígena (ASI), Movimiento Integración Regional (MIR), Movimiento Popular Unido (MPU), and AfroVives (VIVES). The sixth minority is the Movimiento de Apertura Liberal (MOAL, $1.34\%$) with two seats. The Polo Democrático Alternativo (PDA, $3.3\%$) is the only opposition party.

We implement the one--dimensional Bayesian ideal point estimator under the technical details exposed by \cite{luque2022bayesian}. The model provides inferences about the ideal points of 148 parliamentarians (two legislators are anchors for model identification) and about the discrimination and approval parameters of the 560 voting lists, i.e., we estimate 1268 parameters. We eliminate missing data as they have no substantial impact on our findings \citep[see][]{luque2022bayesian}. In future methodological research, roll call votes of this deliberative body could be helpful to illustrate patterns of parliamentary conduct in the region when abstention is frequent (the data set has $44\%$ missing data, of which $39\%$ correspond to abstentions and the remainder to absences).

\begin{figure}[!t]
	\centering
	\includegraphics[scale=0.5]{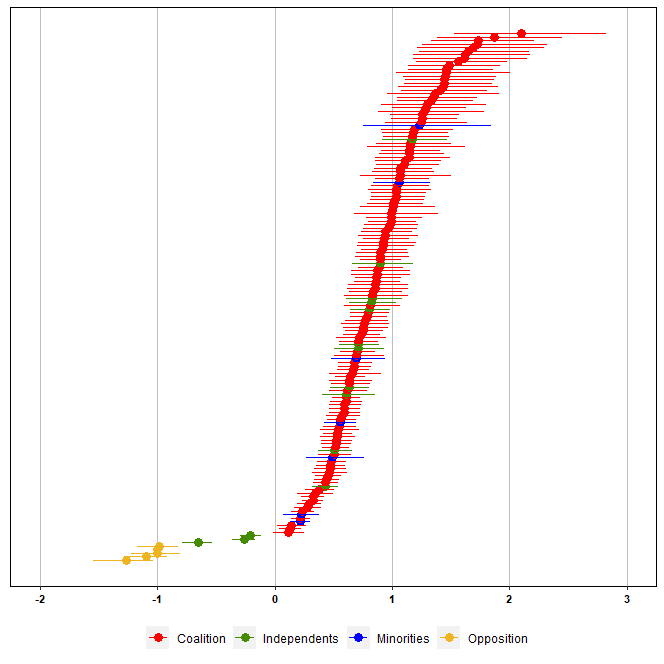}
	\caption{Ideal point estimates along with their 95\% credible bands.}
	\label{fig_fig}
\end{figure}

As shown in Figure \ref{fig_fig}, the ideal point estimates $\beta_i$ vary between $-1.47$ and $2.22$. For all opposition members, the posterior mean of their ideal point is negative and significantly different from zero (i.e., the corresponding credibility interval does not contain zero). In the case of the governing coalition members, the posterior mean of their ideal point is positive, and $99\%$ of the deputies of this political group have statistically significant ideal points. On the other hand, the ideal points of minorities and independent parties are more dispersed throughout the political space, all being significantly different from zero. The distribution of the ideal points in the political space of the lower house is similar to the upper house for the same period \citep[see][]{luque2022bayesian}.

The distribution of the ideal points indicates that the underlying latent feature of the nominal voting of the lower house is not ideological (left--right) because on the same side of the political spectrum are historically antagonistic parties \citep[LC and CC, see][]{mazzuca2009political}. Our findings indicate a latent non--ideological trait that divides the spectrum into opposition--non-opposition since there are political groups that do not openly declare themselves as part of or against the government (e.g., independents and minorities). Then, the recovered dimension reflects power control between hegemonic parties within the legislative body, with some minority and independent members taking a moderate (centrist) position.

On the other hand, we identified that 447 of the 560 discrimination parameters, $\alpha_j$, are significantly different from zero. Then, $79.8\%$ of the motions discriminate between legislators across the political spectrum. Roll calls segregating among deputies are mostly related to security, defense, and public forces ($23.04\%$), social security and health ($19.91\%$), and justice ($20.36\%$). So, we argue that power control within the chamber revolves around these issues. Further, we determined that $85.46\%$ of motions discriminating this political dimension are from a government initiative and also that $95.08\%$ are list passes debated during the first three legislative years.

Furthermore, we note that motions related to the environment, mines and energy, and recreation and sports do not discriminate for the proposed dimension. The 113 motions that do not discriminate in the retrieved policy continuum possibly store information about the second dimension of political space. Finally, we examine the posterior predictive distribution of some test statistics. In none of these cases the posterior predictive $p$-value is an extreme value (less than 0.05 or greater than 0.95). Therefore, the proposed one-dimensional model reveals a good fit for 2010--2014 Colombian House of Representatives nominal vote data set.

\section{Discussion}\label{sec:dis}

The taxonomy proposed in this document specifies substantive themes that inspire research on legislative electoral behavior in different contexts, particularly in Latin America. Future research aims to identify additional lines, phenomena, or substantive hypotheses subject to analysis through the methodologies under review that allows social scientists to extend our categorization.

We identify that four of the seven lines present empirical evidence of legislative electoral behavior in Latin America via the Ideal Bayesian point estimator. Then, in some countries of the region, there are available studies about the political space dimension, the change in the preferences of political actors, the influence of national political leaders or groups, and strategic abstentions. However, to the best of our knowledge, in Latin America, no research uses the Bayesian ideal point estimator to study issues related to voting restricted to the nature of the agenda and the case of extremes voting together. In addition, the available literature only points out to the parliaments of Argentina \citep{jones2005party, rosas2008models}, Brazil \citep{mcdonnell2017formal, tsai2020influence, zucco2011distinguishing}, Colombia \citep{luque2022bayesian}, Paraguay \citep{ribeiro2021legislative}, and Uruguay \citep{zucco2013legislative} as case studies. 
In this sense, Latin American parliaments within a democratic presidential system are exhibited as novel scenarios to operationalize the electoral behavior of legislative bodies through roll call data under a Bayesian approach.

Quantitative studies on parliamentary electoral behavior in Latin America based on the application, adaptation, and generalization of the Bayesian ideal point estimator are emerging as promising research fields. 
In this regard, several aspects are of interest in the framework of the proposed taxonomy. For example, although there are studies on the dimension of the political space, it is still necessary to delve into the 
dynamic and collective nature of the dimensions proposed for the different parliaments \citep[e.g.,][] {moser2019multiple}. Furthermore, the nature of higher dimensions in the region's legislatures is also an object of study. The latter requires specifying the methodological conditions to propose multidimensional spaces in the legislative chambers of Latin America \citep[see][]{jackman2001multidimensional}. In addition, future research can evaluate the variation in the dimension of the political space when votes are segregated by issue, initiative, or other kinds of categorization \citep[e.g.,][]{lofland2017Assessing}.

Regarding the identification of pivotal legislators, future research should provide empirical evidence on the deputies who are essential for the approval of new political alternatives or hinder voting in Latin American parliaments \citep[e.g.,][]{clinton2004statistical}). Furthermore, research framed in voting restricted to the nature of the agenda indicates an opportunity to investigate the importance of parameters of voting alternatives when studying legislative electoral conduct. The latter would allow analysts to know  political scenarios under which there are a greater legislative activity or probability of changing the status quo and identify possible political strategies to establish the legislative agenda in the deliberative bodies of the region. Furthermore, from a technical point of view, more evidence is needed regarding the consistency of the constrained and unconstrained estimators \citep[see][]{clinton2001agenda}.

In the United States, the effect of party switching on legislators' preferences has been extensively studied \citep[see][]{clinton2004statistical}. However, this direction has yet to be studied in Latin America. Additionally, given the characteristics of Latin American parliaments, it is quite important to investigate the variation in the preferences of members of Congress. For example, when there is a change of political groups (e.g., entry and exit of parties to the ruling coalition) or leadership within parliaments (e.g., change of president in the legislative chambers). On the other hand, studies in line with the evolution of legislators' preferences admit the contrast between legislative chambers (e.g., plenary and commissions) to discover changes in the latent features that underlie the vote of deputies. In addition, interested researchers could evaluate the potential of the estimator in its dynamic version to provide empirical evidence on the impact of party fragmentation and the proliferation of political groups on legislative electoral behavior in the region's countries.

Concerning the influence of national political leaders or groups, future research should focus on incorporating multiple incentives  (e.g., party and coalition) into the estimator in the Latin American case. Moreover,
we have yet to review applications of the estimator that incorporate external factors \citep[e.g., media, opinions of the electorate, foreign political groups, or electoral processes, among others; see][]{ribeiro2021legislative, fouirnaies2022electoral}
to examine other determinants of the voting decision of parliamentarians. On the other hand, the line of strategic abstentions traces an alternative to investigating the nature of the dynamics of non-participation in the vote of the deputies in the legislative institutions of Latin America \citep[e.g.,][]{rosas2015no}.

Finally, the line of extremes voting together is a novel methodological alternative to evaluate different technical aspects. \citep[e.g., the geometry underlying political space in Latin American legislatures, see][]{yu2019spherical}. Again, to the best of our knowledge, research in Latin America has yet to explore spherical latent factor voting models for binary and ordinal data. In this document, we do not describe non-parametric implementations of the estimator \citep[see][]{tahk2018nonparametric, shiraito2022non}. Such approaches will be discussed elsewhere.

\bibliography{references}

\begin{thebibliography}{}

\bibitem[Ainsley et~al., 2020]{ainsley2020roll}
Ainsley, C., Carrubba, C.~J., Crisp, B.~F., Demirkaya, B., Gabel, M.~J., and
  Hadzic, D. (2020).
\newblock Roll-call vote selection: Implications for the study of legislative
  politics.
\newblock {\em American Political Science Review}, 114(3):691--706.

\bibitem[Albert and Chib, 1993]{albert1993bayesian}
Albert, J.~H. and Chib, S. (1993).
\newblock Bayesian analysis of binary and polychotomous response data.
\newblock {\em Journal of the American statistical Association},
  88(422):669--679.

\bibitem[Aldrich et~al., 2014]{aldrich2014polarization}
Aldrich, J.~H., Montgomery, J.~M., and Sparks, D.~B. (2014).
\newblock Polarization and ideology: Partisan sources of low dimensionality in
  scaled roll call analyses.
\newblock {\em Political Analysis}, pages 435--456.

\bibitem[Alem{\'a}n, 2008]{aleman2008policy}
Alem{\'a}n, E. (2008).
\newblock Policy positions in the chilean senate: An analysis of coauthorship
  and roll call data.
\newblock {\em Brazilian Political Science Review (Online)}, 3(SE):0--0.

\bibitem[Alem{\'a}n et~al., 2009]{aleman2009comparing}
Alem{\'a}n, E., Calvo, E., Jones, M.~P., and Kaplan, N. (2009).
\newblock Comparing cosponsorship and roll-call ideal points.
\newblock {\em Legislative Studies Quarterly}, 34(1):87--116.

\bibitem[Alem{\'a}n et~al., 2018]{aleman2018disentangling}
Alem{\'a}n, E., Micozzi, J.~P., Pinto, P.~M., and Saiegh, S. (2018).
\newblock Disentangling the role of ideology and partisanship in legislative
  voting: evidence from argentina.
\newblock {\em Legislative Studies Quarterly}, 43(2):245--273.

\bibitem[Amemiya, 1984]{amemiya1984tobit}
Amemiya, T. (1984).
\newblock Tobit models: A survey.
\newblock {\em Journal of econometrics}, 24(1-2):3--61.

\bibitem[Ames, 2002]{ames2002party}
Ames, B. (2002).
\newblock Party discipline in the chamber of deputies.
\newblock {\em Legislative Politics in Latin America}, pages 185--221.

\bibitem[Arrow, 1990]{arrow1990advances}
Arrow, K. (1990).
\newblock {\em Advances in the spatial theory of voting}.
\newblock Cambridge University Press.

\bibitem[Asmussen and Jo, 2016]{asmussen2016anchors}
Asmussen, N. and Jo, J. (2016).
\newblock Anchors away: a new approach for estimating ideal points comparable
  across time and chambers.
\newblock {\em Political Analysis}, pages 172--188.

\bibitem[Aumann, 1964]{aumann1964m}
Aumann, R. (1964).
\newblock The bargaining set for cooperative games.
\newblock {\em Advances in Game Theory}, pages 443--476.

\bibitem[Bailey et~al., 2017]{bailey2017estimating}
Bailey, M.~A., Strezhnev, A., and Voeten, E. (2017).
\newblock Estimating dynamic state preferences from united nations voting data.
\newblock {\em Journal of Conflict Resolution}, 61(2):430--456.

\bibitem[Benoit and Laver, 2012]{benoit2012dimensionality}
Benoit, K. and Laver, M. (2012).
\newblock The dimensionality of political space: Epistemological and
  methodological considerations.
\newblock {\em European Union Politics}, 13(2):194--218.

\bibitem[Bernardo et~al., 2003]{bernardo2003bayesian}
Bernardo, J., Bayarri, M., Berger, J., Dawid, A., Heckerman, D., Smith, A., and
  West, M. (2003).
\newblock Bayesian factor regression models in the “large p, small n”
  paradigm.
\newblock {\em Bayesian statistics}, 7:733--742.

\bibitem[Binding and Stoetzer, 2022]{binding2022non}
Binding, G. and Stoetzer, L.~F. (2022).
\newblock Non-separable preferences in the statistical analysis of roll call
  votes.
\newblock {\em Political Analysis}, pages 1--14.

\bibitem[Black et~al., 1958]{black1958theory}
Black, D. et~al. (1958).
\newblock {\em The theory of committees and elections}.
\newblock Springer.

\bibitem[Brazill and Grofman, 2002]{brazill2002factor}
Brazill, T.~J. and Grofman, B. (2002).
\newblock Factor analysis versus multi-dimensional scaling: binary choice
  roll-call voting and the us supreme court.
\newblock {\em Social Networks}, 24(3):201--229.

\bibitem[Cahoon et~al., 1978]{cahoon1978statistical}
Cahoon, L., Hinich, M.~J., and Ordeshook, P.~C. (1978).
\newblock A statistical multidimensional scaling method based on the spatial
  theory of voting.
\newblock In {\em Graphical representation of multivariate data}, pages
  243--278. Elsevier.

\bibitem[Carey, 1998]{carey1998parties}
Carey, J.~M. (1998).
\newblock {\em Parties, Coalitions, and the Chilean Congress in the 1990s}.
\newblock Latin American Studies Association.

\bibitem[Carlin and Louis, 2008]{carlin2008bayesian}
Carlin, B.~P. and Louis, T.~A. (2008).
\newblock {\em Bayesian methods for data analysis}.
\newblock CRC Press.

\bibitem[Carroll et~al., 2009a]{carroll2009comparing}
Carroll, R., Lewis, J.~B., Lo, J., Poole, K.~T., and Rosenthal, H. (2009a).
\newblock Comparing nominate and ideal: Points of difference and monte carlo
  tests.
\newblock {\em Legislative Studies Quarterly}, 34(4):555--591.

\bibitem[Carroll et~al., 2009b]{carroll2009measuring}
Carroll, R., Lewis, J.~B., Lo, J., Poole, K.~T., and Rosenthal, H. (2009b).
\newblock Measuring bias and uncertainty in dw-nominate ideal point estimates
  via the parametric bootstrap.
\newblock {\em Political analysis}, pages 261--275.

\bibitem[Carroll et~al., 2013]{carroll2013structure}
Carroll, R., Lewis, J.~B., Lo, J., Poole, K.~T., and Rosenthal, H. (2013).
\newblock The structure of utility in spatial models of voting.
\newblock {\em American Journal of Political Science}, 57(4):1008--1028.

\bibitem[Carroll and Pach{\'o}n, 2016]{carroll2016unrealized}
Carroll, R. and Pach{\'o}n, M. (2016).
\newblock The unrealized potential of presidential coalitions in {C}olombia.
\newblock {\em Legislative Institutions and Lawmaking in Latin America}, pages
  122--147.

\bibitem[Carvalho et~al., 2008]{carvalho2008high}
Carvalho, C.~M., Chang, J., Lucas, J.~E., Nevins, J.~R., Wang, Q., and West, M.
  (2008).
\newblock High-dimensional sparse factor modeling: applications in gene
  expression genomics.
\newblock {\em Journal of the American Statistical Association},
  103(484):1438--1456.

\bibitem[Castillo et~al., 2015]{castillo2015bayesian}
Castillo, I., Schmidt-Hieber, J., Van~der Vaart, A., et~al. (2015).
\newblock Bayesian linear regression with sparse priors.
\newblock {\em Annals of Statistics}, 43(5):1986--2018.

\bibitem[Caughey and Schickler, 2016]{caughey2016substance}
Caughey, D. and Schickler, E. (2016).
\newblock Substance and change in congressional ideology: Nominate and its
  alternatives.
\newblock {\em Studies in American Political Development}, 30(2):128--146.

\bibitem[Cheibub et~al., 2009]{cheibub2009political}
Cheibub, J.~A., Figueiredo, A., and Limongi, F. (2009).
\newblock Political parties and governors as determinants of legislative
  behavior in brazil's chamber of deputies, 1988--2006.
\newblock {\em Latin American Politics and Society}, 51(1):1--30.

\bibitem[Chib and Greenberg, 1998]{chib1998analysis}
Chib, S. and Greenberg, E. (1998).
\newblock Analysis of multivariate probit models.
\newblock {\em Biometrika}, 85(2):347--361.

\bibitem[Clerici, 2021]{clerici2021legislative}
Clerici, P. (2021).
\newblock Legislative territorialization: The impact of a decentralized party
  system on individual legislative behavior in argentina.
\newblock {\em Publius: The Journal of Federalism}, 51(1):104--130.

\bibitem[Clinton et~al., 2004]{clinton2004statistical}
Clinton, J., Jackman, S., and Rivers, D. (2004).
\newblock The statistical analysis of roll call data.
\newblock {\em American Political Science Review}, pages 355--370.

\bibitem[Clinton, 2012]{clinton2012using}
Clinton, J.~D. (2012).
\newblock Using roll call estimates to test models of politics.
\newblock {\em Annual Review of Political Science}, 15:79--99.

\bibitem[Clinton and Jackman, 2009]{clinton2009simulate}
Clinton, J.~D. and Jackman, S. (2009).
\newblock To simulate or nominate?
\newblock {\em Legislative Studies Quarterly}, 34(4):593--621.

\bibitem[Clinton and Meirowitz, 2001]{clinton2001agenda}
Clinton, J.~D. and Meirowitz, A. (2001).
\newblock Agenda constrained legislator ideal points and the spatial voting
  model.
\newblock {\em Political Analysis}, pages 242--259.

\bibitem[Cohen and Noll, 1991]{cohen1991vote}
Cohen, L.~R. and Noll, R.~G. (1991).
\newblock How to vote, whether to vote: Strategies for voting and abstaining on
  congressional roll calls.
\newblock {\em Political Behavior}, 13(2):97--127.

\bibitem[Coughlin and Nitzan, 1981]{coughlin1981electoral}
Coughlin, P. and Nitzan, S. (1981).
\newblock Electoral outcomes with probabilistic voting and nash social welfare
  maxima.
\newblock {\em Journal of Public Economics}, 15(1):113--121.

\bibitem[Davis and Hinich, 1965]{davis1965mathematical}
Davis, O.~A. and Hinich, M.~J. (1965).
\newblock {\em A mathematical model of policy formation in a democratic
  society}.
\newblock Graduate School of Industrial Administration, Carnegie Institute of
  Technology.

\bibitem[Davis et~al., 1970]{davis1970expository}
Davis, O.~A., Hinich, M.~J., and Ordeshook, P.~C. (1970).
\newblock An expository development of a mathematical model of the electoral
  process.
\newblock {\em The American Political Science Review}, 64(2):426--448.

\bibitem[De~Leeuw, 2006]{de2006principal}
De~Leeuw, J. (2006).
\newblock Principal component analysis of binary data by iterated singular
  value decomposition.
\newblock {\em Computational statistics \& data analysis}, 50(1):21--39.

\bibitem[De~Vries and Marks, 2012]{de2012struggle}
De~Vries, C.~E. and Marks, G. (2012).
\newblock The struggle over dimensionality: A note on theory and empirics.
\newblock {\em European Union Politics}, 13(2):185--193.

\bibitem[Dougherty et~al., 2014]{dougherty2014partisan}
Dougherty, K.~L., Lynch, M.~S., and Madonna, A.~J. (2014).
\newblock Partisan agenda control and the dimensionality of congress.
\newblock {\em American Politics Research}, 42(4):600--627.

\bibitem[Downs, 1957]{downs1957economic}
Downs, A. (1957).
\newblock An economic theory of political action in a democracy.
\newblock {\em Journal of political economy}, 65(2):135--150.

\bibitem[Enelow and Hinich, 1984]{enelow1984spatial}
Enelow, J.~M. and Hinich, M.~J. (1984).
\newblock {\em The spatial theory of voting: An introduction}.
\newblock CUP Archive.

\bibitem[Figueiredo and Limongi, 2000]{figueiredo2000presidential}
Figueiredo, A.~C. and Limongi, F. (2000).
\newblock Presidential power, legislative organization, and party behavior in
  brazil.
\newblock {\em Comparative Politics}, pages 151--170.

\bibitem[Fouirnaies and Hall, 2022]{fouirnaies2022electoral}
Fouirnaies, A. and Hall, A.~B. (2022).
\newblock How do electoral incentives affect legislator behavior? evidence from
  us state legislatures.
\newblock {\em American Political Science Review}, 116(2):662--676.

\bibitem[Gamerman and Lopes, 2006]{gamerman2006markov}
Gamerman, D. and Lopes, H. (2006).
\newblock {\em Markov chain Monte Carlo: stochastic simulation for Bayesian
  inference}.
\newblock CRC Press.

\bibitem[Gamm and Huber, 2002]{gamm2002legislatures}
Gamm, G. and Huber, J. (2002).
\newblock Legislatures as political institutions: Beyond the contemporary
  congress.
\newblock In IraKatznelson and Milner, H.~V., editors, {\em Political science:
  State of the discipline}, pages 313--341. New York: W.W. Norton.

\bibitem[Grier et~al., 2022]{grier2022campaign}
Grier, K., Grier, R., and Mkrtchian, G. (2022).
\newblock Campaign contributions and roll-call voting in the us house of
  representatives: The case of the sugar industry.
\newblock {\em American Political Science Review}, pages 1--7.

\bibitem[Hahn and Soyer, 2005]{hahn2005probit}
Hahn, E.~D. and Soyer, R. (2005).
\newblock Probit and logit models: Differences in the multivariate realm.
\newblock {\em The Journal of the Royal Statistical Society, Series B}, pages
  1--12.

\bibitem[Hahn et~al., 2012]{richard2012sparse}
Hahn, R.~P., Carvalho, C.~M., and Scott, J.~G. (2012).
\newblock A sparse factor analytic probit model for congressional voting
  patterns.
\newblock {\em Journal of the Royal Statistical Society: Series C (Applied
  Statistics)}, 61(4):619--635.

\bibitem[Hansford et~al., 2022]{hansford2022estimating}
Hansford, T.~G., Depaoli, S., and Canelo, K.~S. (2022).
\newblock Estimating the ideal points of organized interests in legal policy
  space.
\newblock {\em Justice System Journal}, pages 1--12.

\bibitem[Hinich and Munger, 1996]{hinich1996ideology}
Hinich, M.~J. and Munger, M.~C. (1996).
\newblock {\em Ideology and the theory of political choice}.
\newblock University of Michigan Press.

\bibitem[Hinich et~al., 1997]{hinich1997analytical}
Hinich, M.~J., Munger, M.~C., et~al. (1997).
\newblock {\em Analytical politics}.
\newblock Cambridge university press.

\bibitem[Hinich and Ordeshook, 1970]{hinich1970plurality}
Hinich, M.~J. and Ordeshook, P.~C. (1970).
\newblock Plurality maximization vs vote maximization: A spatial analysis with
  variable participation.
\newblock {\em The American Political Science Review}, 64(3):772--791.

\bibitem[Hix et~al., 2005]{hix2005power}
Hix, S., Noury, A., and Roland, G. (2005).
\newblock Power to the parties: cohesion and competition in the european
  parliament 1979-2001.
\newblock {\em British Journal of Political Science}, pages 209--234.

\bibitem[Jackman, 2001]{jackman2001multidimensional}
Jackman, S. (2001).
\newblock Multidimensional analysis of roll call data via bayesian simulation:
  Identification, estimation, inference, and model checking.
\newblock {\em Political Analysis}, 9(3):227--241.

\bibitem[Jackman, 2004]{jackman2004bayesian}
Jackman, S. (2004).
\newblock Bayesian analysis for political research.
\newblock {\em Annu. Rev. Polit. Sci.}, 7:483--505.

\bibitem[Jackman, 2009]{jackman2009bayesian}
Jackman, S. (2009).
\newblock {\em Bayesian analysis for the social sciences}, volume 846.
\newblock John Wiley \& Sons.

\bibitem[Jones and Hwang, 2005a]{jones2005party}
Jones, M.~P. and Hwang, W. (2005a).
\newblock Party government in presidential democracies: Extending cartel theory
  beyond the us congress.
\newblock {\em American Journal of Political Science}, 49(2):267--282.

\bibitem[Jones and Hwang, 2005b]{jones2005provincial}
Jones, M.~P. and Hwang, W. (2005b).
\newblock {\em Provincial party bosses: Keystone of the Argentine Congress},
  pages 115--138.
\newblock Pennsylvania State University Press University Park.

\bibitem[Jones et~al., 2009]{jones2009government}
Jones, M.~P., Hwang, W., and Micozzi, J.~P. (2009).
\newblock Government and opposition in the argentine congress, 1989-2007:
  Understanding inter-party dynamics through roll call vote analysis.
\newblock {\em Journal of Politics in Latin America}, 1(1):67--96.

\bibitem[Kass and Wasserman, 1996]{kass1996selection}
Kass, R.~E. and Wasserman, L. (1996).
\newblock The selection of prior distributions by formal rules.
\newblock {\em Journal of the American statistical Association},
  91(435):1343--1370.

\bibitem[Kim et~al., 2018]{kim2018estimating}
Kim, I.~S., Londregan, J., and Ratkovic, M. (2018).
\newblock Estimating spatial preferences from votes and text.
\newblock {\em Political Analysis}, 26(2):210--229.

\bibitem[Krehbiel, 1988]{krehbiel1988spatial}
Krehbiel, K. (1988).
\newblock Spatial models of legislative choice.
\newblock {\em Legislative Studies Quarterly}, pages 259--319.

\bibitem[Krehbiel, 1998]{krehbiel1998pivotal}
Krehbiel, K. (1998).
\newblock {\em Pivotal politics: A theory of US lawmaking}.
\newblock University of Chicago Press.

\bibitem[Kromer, 2005]{kromer2005determinants}
Kromer, M.~K. (2005).
\newblock Determinants of abstention in the united states house of
  representatives: an analysis of the 102nd through the 107th sessions.
\newblock Master's thesis, Louisiana State University, Baton Rouge, LA.

\bibitem[Lauderdale and Clark, 2014]{lauderdale2014scaling}
Lauderdale, B.~E. and Clark, T.~S. (2014).
\newblock Scaling politically meaningful dimensions using texts and votes.
\newblock {\em American Journal of Political Science}, 58(3):754--771.

\bibitem[Lee, 2018]{lee2018introduction}
Lee, J.~M. (2018).
\newblock {\em Introduction to Riemannian manifolds}.
\newblock Springer.

\bibitem[Lewis and Poole, 2004]{lewis2004measuring}
Lewis, J.~B. and Poole, K.~T. (2004).
\newblock Measuring bias and uncertainty in ideal point estimates via the
  parametric bootstrap.
\newblock {\em Political Analysis}, pages 105--127.

\bibitem[Lofland et~al., 2017]{lofland2017Assessing}
Lofland, C.~L., Rodr{\'\i}guez, A., Moser, S., et~al. (2017).
\newblock Assessing differences in legislators’ revealed preferences: A case
  study on the 107th us senate.
\newblock {\em The Annals of Applied Statistics}, 11(1):456--479.

\bibitem[Luque, 2021]{luque2021metodos}
Luque, C. (2021).
\newblock M{\'e}todos bayesianos para caracterizar el comportamiento
  legislativo del senado colombiano en el periodo 2010-2014.
\newblock Master's thesis, Universidad Santo Tom{\'a}s.

\bibitem[Luque and Sosa, 2022]{luque2022bayesian}
Luque, C. and Sosa, J. (2022).
\newblock A bayesian spatial voting model to characterize the legislative
  behavior of the colombian senate 2010--2014.
\newblock {\em Journal of Applied Statistics}, pages 1--22.

\bibitem[MacRae, 1952]{macrae1952relation}
MacRae, D. (1952).
\newblock The relation between roll call votes and constituencies in the
  massachusetts house of representatives.
\newblock {\em American Political Science Review}, 46(4):1046--1055.

\bibitem[MacRae, 1958]{macrae1958dimensions}
MacRae, D. (1958).
\newblock Dimensions of congressional voting: A statistical study of the house
  of representatives in the eighty-first congress.
\newblock {\em The Journal of Politics}, 1(3).

\bibitem[MacRae, 1965]{macrae1965method}
MacRae, D. (1965).
\newblock A method for identifying issues and factions from legislative votes.
\newblock {\em The American Political science Review}, 59(4):909--926.

\bibitem[Manski, 1977]{manski1977structure}
Manski, C.~F. (1977).
\newblock The structure of random utility models.
\newblock {\em Theory and decision}, 8(3):229.

\bibitem[Martin and Quinn, 2002]{martin2002dynamic}
Martin, A.~D. and Quinn, K.~M. (2002).
\newblock Dynamic ideal point estimation via markov chain monte carlo for the
  us supreme court, 1953--1999.
\newblock {\em Political analysis}, 10(2):134--153.

\bibitem[Mayhew, 1974]{mayhew1974congress}
Mayhew, D.~R. (1974).
\newblock {\em Congress: The electoral connection}.
\newblock Yale university press.

\bibitem[Mazzuca and Robinson, 2009]{mazzuca2009political}
Mazzuca, S. and Robinson, J.~A. (2009).
\newblock Political conflict and power sharing in the origins of modern
  colombia.
\newblock {\em Hispanic American Historical Review}, 89(2):285--321.

\bibitem[McCarty et~al., 2001]{mccarty2001hunt}
McCarty, N., Poole, K.~T., and Rosenthal, H. (2001).
\newblock The hunt for party discipline in congress.
\newblock {\em American Political Science Review}, pages 673--687.

\bibitem[McCullagh, 2018]{mccullagh2018generalized}
McCullagh, P. (2018).
\newblock {\em Generalized linear models}.
\newblock Routledge.

\bibitem[McDonnell, 2017]{mcdonnell2017formal}
McDonnell, R.~M. (2017).
\newblock Formal comparisons of legislative institutions: Ideal points from
  brazilian legislatures.
\newblock {\em Brazilian Political Science Review}, 11(1).

\bibitem[McDonnell et~al., 2019]{mcdonnell2019congressbr}
McDonnell, R.~M., Duarte, G.~J., and Freire, D. (2019).
\newblock congressbr: An r package for analyzing data from brazil’s chamber
  of deputies and federal senate.
\newblock {\em Latin American Research Review}, 54(4).

\bibitem[McFadden, 1976]{mcfadden1976quantal}
McFadden, D.~L. (1976).
\newblock Quantal choice analaysis: A survey.
\newblock In {\em Annals of Economic and Social Measurement, Volume 5, number
  4}, pages 363--390. NBER.

\bibitem[McKelvey et~al., 1978]{mckelvey1978competitive}
McKelvey, R.~D., Ordeshook, P.~C., and Winer, M.~D. (1978).
\newblock The competitive solution for n-person games without transferable
  utility, with an application to committee games.
\newblock {\em American Political Science Review}, 72(2):599--615.

\bibitem[Miller, 1956]{miller1956magical}
Miller, G.~A. (1956).
\newblock The magical number seven, plus or minus two: Some limits on our
  capacity for processing information.
\newblock {\em Psychological review}, 63(2):81.

\bibitem[Morales, 2021]{morales2021legislating}
Morales, J.~S. (2021).
\newblock Legislating during war: Conflict and politics in colombia.
\newblock {\em Journal of Public Economics}, 193:104325.

\bibitem[Morgenstern, 2003]{morgenstern2003patterns}
Morgenstern, S. (2003).
\newblock {\em Patterns of legislative politics: roll-call voting in Latin
  America and the United States}.
\newblock Cambridge University Press.

\bibitem[Moser et~al., 2021]{moser2019multiple}
Moser, S., Rodr{\'\i}guez, A., and Lofland, C.~L. (2021).
\newblock Multiple ideal points: Revealed preferences in different domains.
\newblock {\em Political Analysis}, 29(2):139--166.

\bibitem[Murray et~al., 2013]{murray2013bayesian}
Murray, J.~S., Dunson, D.~B., Carin, L., and Lucas, J.~E. (2013).
\newblock Bayesian gaussian copula factor models for mixed data.
\newblock {\em Journal of the American Statistical Association},
  108(502):656--665.

\bibitem[Neto, 2002]{neto2002presidential}
Neto, O.~A. (2002).
\newblock Presidential cabinets, electoral cycles, and coalition discipline in
  brazil.
\newblock {\em Legislative Politics in Latin America}, pages 48--78.

\bibitem[Onuki et~al., 2009]{onuki2009political}
Onuki, J., Ribeiro, P.~F., and Oliveira, A. J.~d. (2009).
\newblock Political parties, foreign policy and ideology: Argentina and chile
  in comparative perspective.
\newblock {\em Brazilian Political Science Review (Online)}, 4(SE):0--0.

\bibitem[Ormerod and Wand, 2010]{ormerod2010explaining}
Ormerod, J.~T. and Wand, M.~P. (2010).
\newblock Explaining variational approximations.
\newblock {\em The American Statistician}, 64(2):140--153.

\bibitem[Pach{\'o}n and Johnson, 2016]{pachon2016s}
Pach{\'o}n, M. and Johnson, G.~B. (2016).
\newblock When's the party (or coalition)? agenda-setting in a highly
  fragmented, decentralized legislature.
\newblock {\em Journal of Politics in Latin America}, 8(2):71--100.

\bibitem[Pati et~al., 2014]{pati2014posterior}
Pati, D., Bhattacharya, A., Pillai, N.~S., Dunson, D., et~al. (2014).
\newblock Posterior contraction in sparse bayesian factor models for massive
  covariance matrices.
\newblock {\em Annals of Statistics}, 42(3):1102--1130.

\bibitem[Pereira and Mueller, 2004a]{pereira2004cost}
Pereira, C. and Mueller, B. (2004a).
\newblock The cost of governing: Strategic behavior of the president and
  legislators in brazil’s budgetary process.
\newblock {\em Comparative Political Studies}, 37(7):781--815.

\bibitem[Pereira and Mueller, 2004b]{pereira2004theory}
Pereira, C. and Mueller, B. (2004b).
\newblock A theory of executive dominance of congressional politics: the
  committee system in the brazilian chamber of deputies.
\newblock {\em The Journal of Legislative Studies}, 10(1):9--49.

\bibitem[Poole, 2005]{poole2005spatial}
Poole, K.~T. (2005).
\newblock {\em Spatial models of parliamentary voting}.
\newblock Cambridge University Press.

\bibitem[Poole, 2007]{poole2007changing}
Poole, K.~T. (2007).
\newblock Changing minds? not in congress!
\newblock {\em Public Choice}, 131(3-4):435--451.

\bibitem[Poole and Rosenthal, 1984]{poole1984us}
Poole, K.~T. and Rosenthal, H. (1984).
\newblock Us presidential elections 1968-80: A spatial analysis.
\newblock {\em American Journal of Political Science}, pages 282--312.

\bibitem[Poole and Rosenthal, 1985]{poole1985spatial}
Poole, K.~T. and Rosenthal, H. (1985).
\newblock A spatial model for legislative roll call analysis.
\newblock {\em American Journal of Political Science}, pages 357--384.

\bibitem[Poole and Rosenthal, 1987]{poole1987analysis}
Poole, K.~T. and Rosenthal, H. (1987).
\newblock Analysis of congressional coalition patterns: A unidimensional
  spatial model.
\newblock {\em Legislative Studies Quarterly}, pages 55--75.

\bibitem[Poole and Rosenthal, 2001]{poole2001d}
Poole, K.~T. and Rosenthal, H. (2001).
\newblock D-nominate after 10 years: A comparative update to congress: A
  political-economic history of roll-call voting.
\newblock {\em Legislative Studies Quarterly}, pages 5--29.

\bibitem[Potoski and Talbert, 2000]{potoski2000dimensional}
Potoski, M. and Talbert, J. (2000).
\newblock The dimensional structure of policy outputs: Distributive policy and
  roll call voting.
\newblock {\em Political Research Quarterly}, 53(4):695--710.

\bibitem[Quinn, 2004]{quinn2004bayesian}
Quinn, K.~M. (2004).
\newblock Bayesian factor analysis for mixed ordinal and continuous responses.
\newblock {\em Political Analysis}, 12(4):338--353.

\bibitem[Rasmussen, 2022]{rasmussen2022farmers}
Rasmussen, M.~B. (2022).
\newblock Farmers and the origin of the welfare state: Evidence from 308 roll
  call votes between 1882 and 1940.
\newblock {\em Scandinavian Political Studies}, 45(2):202--226.

\bibitem[Ribeiro et~al., 2021]{ribeiro2021legislative}
Ribeiro, P.~F., Burian, C.~L., and Urdinez, F. (2021).
\newblock Legislative behavior, mass media, and foreign policy making: The case
  of paraguay.
\newblock {\em Latin American Research Review}, 56(2).

\bibitem[Rivers, 2003]{rivers2003identification}
Rivers, D. (2003).
\newblock Identification of multidimensional item-response models.
\newblock {\em Typescript. Department of Political Science, Stanford
  University}.

\bibitem[Roberts, 2007]{roberts2007statistical}
Roberts, J.~M. (2007).
\newblock The statistical analysis of roll-call data: A cautionary tale.
\newblock {\em Legislative Studies Quarterly}, 32(3):341--360.

\bibitem[Roberts et~al., 2016]{roberts2016dimensionality}
Roberts, J.~M., Smith, S.~S., and Haptonstahl, S.~R. (2016).
\newblock The dimensionality of congressional voting reconsidered.
\newblock {\em American Politics Research}, 44(5):794--815.

\bibitem[Rodr{\'\i}guez and Moser, 2015]{rodriguez2015measuring}
Rodr{\'\i}guez, A. and Moser, S. (2015).
\newblock Measuring and accounting for strategic abstentions in the us senate,
  1989--2012.
\newblock {\em Journal of the Royal Statistical Society: Series C: Applied
  Statistics}, pages 779--797.

\bibitem[Romer and Rosenthal, 1978]{romer1978political}
Romer, T. and Rosenthal, H. (1978).
\newblock Political resource allocation, controlled agendas, and the status
  quo.
\newblock {\em Public choice}, 33(4):27--43.

\bibitem[Rosas, 2005]{rosas2005ideological}
Rosas, G. (2005).
\newblock The ideological organization of latin american legislative parties:
  An empirical analysis of elite policy preferences.
\newblock {\em Comparative Political Studies}, 38(7):824--849.

\bibitem[Rosas and Shomer, 2008]{rosas2008models}
Rosas, G. and Shomer, Y. (2008).
\newblock Models of nonresponse in legislative politics.
\newblock {\em Legislative Studies Quarterly}, 33(4):573--601.

\bibitem[Rosas et~al., 2015]{rosas2015no}
Rosas, G., Shomer, Y., and Haptonstahl, S.~R. (2015).
\newblock No news is news: Nonignorable nonresponse in roll-call data analysis.
\newblock {\em American Journal of Political Science}, 59(2):511--528.

\bibitem[Scott and Berger, 2006]{scott2006exploration}
Scott, J.~G. and Berger, J.~O. (2006).
\newblock An exploration of aspects of bayesian multiple testing.
\newblock {\em Journal of statistical planning and inference},
  136(7):2144--2162.

\bibitem[Scott and Berger, 2010]{scott2010bayes}
Scott, J.~G. and Berger, J.~O. (2010).
\newblock Bayes and empirical-bayes multiplicity adjustment in the
  variable-selection problem.
\newblock {\em The Annals of Statistics}, pages 2587--2619.

\bibitem[Seabra and Mesquita, 2022]{seabra2022beyond}
Seabra, P. and Mesquita, R. (2022).
\newblock Beyond roll-call voting: Sponsorship dynamics at the un general
  assembly.
\newblock {\em International Studies Quarterly}, 66(2):sqac008.

\bibitem[Shepsle, 1979]{shepsle1979institutional}
Shepsle, K.~A. (1979).
\newblock Institutional arrangements and equilibrium in multidimensional voting
  models.
\newblock {\em American Journal of Political Science}, pages 27--59.

\bibitem[Shiraito et~al., 2022]{shiraito2022non}
Shiraito, Y., Lo, J., and Olivella, S. (2022).
\newblock A non-parametric bayesian model for detecting differential item
  functioning: An application to political representation in the us.
\newblock {\em arXiv preprint arXiv:2205.05934}.

\bibitem[Shor et~al., 2010]{shor2010bridge}
Shor, B., Berry, C., and McCarty, N. (2010).
\newblock A bridge to somewhere: Mapping state and congressional ideology on a
  cross-institutional common space.
\newblock {\em Legislative Studies Quarterly}, 35(3):417--448.

\bibitem[Shor and McCarty, 2011]{shor2011ideological}
Shor, B. and McCarty, N. (2011).
\newblock The ideological mapping of american legislatures.
\newblock {\em American Political Science Review}, pages 530--551.

\bibitem[Snyder~Jr and Groseclose, 2000]{snyder2000estimating}
Snyder~Jr, J.~M. and Groseclose, T. (2000).
\newblock Estimating party influence in congressional roll-call voting.
\newblock {\em American Journal of Political Science}, pages 193--211.

\bibitem[Sosa and Betancourt, 2022]{sosa2022latent}
Sosa, J. and Betancourt, B. (2022).
\newblock A latent space model for multilayer network data.
\newblock {\em Computational Statistics \& Data Analysis}, page 107432.

\bibitem[Tahk, 2018]{tahk2018nonparametric}
Tahk, A. (2018).
\newblock Nonparametric ideal-point estimation and inference.
\newblock {\em Political Analysis}, 26(2):131--146.

\bibitem[Tajfel, 1981]{tajfel1981human}
Tajfel, H. (1981).
\newblock {\em Human groups and social categories: Studies in social
  psychology}.
\newblock Cup Archive.

\bibitem[Talbert and Potoski, 2002]{talbert2002setting}
Talbert, J.~C. and Potoski, M. (2002).
\newblock Setting the legislative agenda: The dimensional structure of bill
  cosponsoring and floor voting.
\newblock {\em Journal of Politics}, 64(3):864--891.

\bibitem[Thurner, 2000]{thurner2000empirical}
Thurner, P.~W. (2000).
\newblock The empirical application of the spatial theory of voting in
  multiparty systems with random utility models.
\newblock {\em Electoral Studies}, 19(4):493--517.

\bibitem[Tiemann, 2019]{tiemann2019shape}
Tiemann, G. (2019).
\newblock The shape of utility functions and voter attitudes towards risk.
\newblock {\em Electoral Studies}, 61:102051.

\bibitem[Treier and Jackman, 2008]{treier2008democracy}
Treier, S. and Jackman, S. (2008).
\newblock Democracy as a latent variable.
\newblock {\em American Journal of Political Science}, 52(1):201--217.

\bibitem[Tsai, 2020]{tsai2020influence}
Tsai, T. (2020).
\newblock The influence of the president and government coalition on roll-call
  voting in brazil, 2003--2006.
\newblock {\em Political Studies Review}, page 1478929920904588.

\bibitem[VanDoren, 1990]{vandoren1990can}
VanDoren, P.~M. (1990).
\newblock Can we learn the causes of congressional decisions from roll-call
  data?
\newblock {\em Legislative Studies Quarterly}, pages 311--340.

\bibitem[Voeten, 2000]{voeten2000clashes}
Voeten, E. (2000).
\newblock Clashes in the assembly.
\newblock {\em International organization}, pages 185--215.

\bibitem[Voeten, 2013]{voeten2013data}
Voeten, E. (2013).
\newblock Data and analyses of voting in the united nations general assembly.
\newblock In {\em Routledge handbook of international organization}, pages
  80--92. Routledge.

\bibitem[Wainer, 1993]{wainer1993differential}
Wainer, P. W. H.~H. (1993).
\newblock {\em Differential item functioning}.
\newblock Psychology Press.

\bibitem[Weisberg and Rusk, 1970]{weisberg1970dimensions}
Weisberg, H.~F. and Rusk, J.~G. (1970).
\newblock Dimensions of candidate evaluation.
\newblock {\em The American Political Science Review}, 64(4):1167--1185.

\bibitem[West and Harrison, 2006]{west2006bayesian}
West, M. and Harrison, J. (2006).
\newblock {\em Bayesian forecasting and dynamic models}.
\newblock Springer Science \& Business Media.

\bibitem[Wolters, 1978]{wolters1978models}
Wolters, M. (1978).
\newblock Models of roll-call behavior.
\newblock {\em Political Methodology}, pages 7--54.

\bibitem[Yu, 2020]{yu2020spherical}
Yu, X. (2020).
\newblock {\em Spherical Latent Factor Model for Binary and Ordinal Data}.
\newblock PhD thesis, UC Santa Cruz.

\bibitem[Yu and Rodriguez, 2019]{yu2019spherical}
Yu, X. and Rodriguez, A. (2019).
\newblock Spherical latent factor model.
\newblock {\em Available at SSRN 3381925}.

\bibitem[Yu and Rodr{\'\i}guez, 2021]{yu2021spatial}
Yu, X. and Rodr{\'\i}guez, A. (2021).
\newblock Spatial voting models in circular spaces: A case study of the us
  house of representatives.
\newblock {\em The Annals of Applied Statistics}, 15(4):1897--1922.

\bibitem[Zellner, 1986]{zellner1986assessing}
Zellner, A. (1986).
\newblock On assessing prior distributions and bayesian regression analysis
  with g-prior distributions.
\newblock {\em Bayesian inference and decision techniques}.

\bibitem[Zucco, 2009]{zucco2009ideology}
Zucco, C. (2009).
\newblock Ideology or what? legislative behavior in multiparty presidential
  settings.
\newblock {\em The Journal of Politics}, 71(3):1076--1092.

\bibitem[Zucco, 2013]{zucco2013legislative}
Zucco, C. (2013).
\newblock Legislative coalitions in presidential systems: the case of uruguay.
\newblock {\em Latin American politics and society}, 55(1):96--118.

\bibitem[Zucco and Lauderdale, 2011]{zucco2011distinguishing}
Zucco, C. and Lauderdale, B.~E. (2011).
\newblock Distinguishing between influences on brazilian legislative behavior.
\newblock {\em Legislative Studies Quarterly}, 36(3):363--396.

\end{thebibliography}
\bibliographystyle{apalike}

\end{document}